\documentclass[useAMS,usenatbib,usegraphicx]{mn2e}

\usepackage{longtable}
\usepackage{setspace}

\newcommand{\arcdeg}{$^{\circ}$}

\newcommand{\maxi}{\mbox{MAXI\,J1659$-$152~}}
\newcommand{\maxinos}{\mbox{MAXI\,J1659$-$152}}

\title[Evolution of the jet and disk of \maxinos]{\vspace{-1.0cm}\begin{spacing}{0.9}{\huge Broadband monitoring tracing the evolution of the jet and disk in the black hole candidate X-ray binary \maxi}\end{spacing}\vspace{-0.5cm}}

\author[A.J. van der Horst et al.]{\parbox{\textwidth}{\begin{spacing}{0.9}{\Large A.J. van~der~Horst$^{1}$\thanks{e-mail: A.J.vanderHorst@uva.nl}, P.A. Curran$^{2}$, J.C.A. Miller-Jones$^{2}$, J.D. Linford$^{3}$, J. Gorosabel$^{4,5,6}$, D.M. Russell$^{7,8}$, A. de~Ugarte~Postigo$^{4,9}$, A.A. Lundgren$^{10}$, G.B. Taylor$^{3}$, D. Maitra$^{11}$, S. Guziy$^{12}$, T.M. Belloni$^{13}$, C. Kouveliotou$^{14}$, P.G. Jonker$^{15,16,17}$, A. Kamble$^{16}$, Z. Paragi$^{18}$, J. Homan$^{19}$, E. Kuulkers$^{20}$, J. Granot$^{21}$, D. Altamirano$^{1}$, M.M. Buxton$^{22}$, A. Castro-Tirado$^{4}$, R.P. Fender$^{23}$, M.A. Garrett$^{24,25}$, N. Gehrels$^{26}$, D.H. Hartmann$^{27}$, J.A. Kennea$^{28}$, H.A. Krimm$^{26,29}$, V. Mangano$^{30}$, E. Ramirez-Ruiz$^{31}$, P. Romano$^{30}$, R.A.M.J. Wijers$^{1}$, R. Wijnands$^{1}$, Y.J. Yang$^{1}$}\end{spacing}}\vspace{0.5cm} \\
\parbox{\textwidth}{\begin{spacing}{0.95}{\small 
$^{1}$Astronomical Institute, University of Amsterdam, Science Park 904, 1098 XH Amsterdam, The Netherlands \\
$^{2}$International Centre for Radio Astronomy Research $-$ Curtin University, GPO Box U1987, Perth, WA 6845, Australia \\
$^{3}$Department of Physics and Astronomy, University of New Mexico, MSC074220, Albuquerque, NM 87131-0001, USA \\
$^{4}$Instituto de Astrof\'{\i}sica de Andaluc\'{\i}a (IAA-CSIC), Glorieta de la Astronom\'{\i}a s/n, 18008, Granada, Spain \\
$^{5}$Unidad Asociada Grupo Ciencia Planetarias UPV/EHU-IAA/CSIC, Departamento de F\'{\i}sica Aplicada I, E.T.S. Ingenier\'{\i}a, Universidad del Pa\'{\i}s Vasco UPV/EHU, Alameda de Urquijo s/n, E-48013 Bilbao, Spain \\
$^{6}$Ikerbasque, Basque Foundation for Science, Alameda de Urquijo 36-5, E-48008 Bilbao, Spain \\
$^{7}$Instituto de Astrof\'{\i}sica de Canarias (IAC), V\'{\i}a L\'{a}ctea s/n, La Laguna E-38205 S/C de Tenerife, Spain \\
$^{8}$Departamento de Astrof\'{\i}sica, Universidad de La Laguna, La Laguna E-38205 S/C de Tenerife, Spain \\
$^{9}$Dark Cosmology Centre, Niels Bohr Institute, Juliane Maries Vej 30, Copenhagen \O, D-2100, Denmark \\
$^{10}$Joint ALMA Observatory, Alonso de C\'{o}rdova 3107, Vitacura - Santiago, Chile \\
$^{11}$Department of Astronomy, University of Michigan, 500 Church Street, Ann Arbor, MI 48109, USA \\
$^{12}$Nikolaev National University, Nikolska 24, Nikolaev, 54030, Ukraine \\
$^{13}$INAF-Osservatorio Astronomico di Brera, Via E. Bianchi 46, I-23807 Merate (LC), Italy \\
$^{14}$Space Science Office, ZP12, NASA/Marshall Space Flight Center, Huntsville, AL 35812, USA \\
$^{15}$SRON, Netherlands Institute for Space Research, Sorbonnelaan 2, 3584CA, Utrecht, The Netherlands \\
$^{16}$Harvard-Smithsonian Center for Astrophysics, 60 Garden Street, Cambridge, MA 02138, USA \\
$^{17}$Department of Astrophysics/IMAPP, Radboud University Nijmegen, P.O. Box 9010, 6500 GL, Nijmegen, The Netherlands \\
$^{18}$Joint Institute for VLBI in Europe, Postbus 2, 7990 AA Dwingeloo, The Netherlands \\
$^{19}$Massachusetts Institute of Technology, Kavli Institute for Astrophysics and Space Research, Cambridge, MA 02139, USA \\
$^{20}$European Space Astronomy Centre (ESA/ESAC), Science Operations Department, 28691 Villanuevadela Ca\~{n}ada (Madrid), Spain \\
$^{21}$Departement of Natural Sciences, The Open University of Israel, P.O. Box 808, Ra'anana 43537, Israel \\
$^{22}$Astronomy Department, Yale University, P.O. Box 208101, New Haven, CT 06520-8101, USA \\
$^{23}$School of Physics and Astronomy, University of Southampton, Southampton, Hampshire SO17 1BJ, UK \\
$^{24}$Netherlands Institute for Radio Astronomy (ASTRON), Postbus 2, 7990 AA Dwingeloo, The Netherlands \\
$^{25}$Leiden Observatory, University of Leiden, Postbus 9513, Leiden 2300 RA, The Netherlands \\
$^{26}$NASA/Goddard Space Flight Center, 8800 Greenbelt Road, Greenbelt, MD 20771, USA \\
$^{27}$Department of Physics and Astronomy, Clemson University, Clemson, SC 29634-0978, USA \\
$^{28}$Department of Astronomy and Astrophysics, Pennsylvania State University, University Park, PA 16802, USA \\
$^{29}$Universities Space Research Association, 10211 Wincopin Circle, Suite 500, Columbia, MD 21044, USA \\
$^{30}$INAF, Istituto di Astrofysica Spaziale e Fisica Cosmica - Palermo, Via U. La Malfa 153, I-90146 Palermo, Italy \\
$^{31}$Department of Astronomy and Astrophysics, University of California, Santa Cruz, CA 95064, USA
}\end{spacing}}\vspace{-0.7cm}}

\begin{document}

\maketitle

\label{firstpage}

\begin{abstract} 
\maxi was discovered on 2010 September 25 as a new X-ray transient, initially identified as a gamma-ray burst, but was later shown to be a new X-ray binary with a black hole as the most likely compact object. 
Dips in the X-ray light curves have revealed that \maxi is the shortest period black hole candidate identified to date. 
Here we present the results of a large observing campaign at radio, sub-millimeter, near-infrared (nIR), optical and ultraviolet (UV) wavelengths. 
We have combined this very rich data set with the available X-ray observations to compile a broadband picture of the evolution of this outburst. 
We have performed broadband spectral modeling, demonstrating the presence of a spectral break at radio frequencies and a relationship between the radio spectrum and X-ray states. 
Also, we have determined physical parameters of the accretion disk and put them into context with respect to the other parameters of the binary system. 
Finally, we have investigated the radio-X-ray and nIR/optical/UV-X-ray correlations up to $\sim3$~years after the outburst onset to examine the link between the jet and the accretion disk, and found that there is no significant jet contribution to the nIR emission when the source is in the soft or intermediate X-ray spectral state, consistent with our detection of the jet break at radio frequencies during these states.
\end{abstract}


\section{Introduction} \label{section:intro}

Black hole X-ray binaries (BHXBs) are usually discovered as transient sources, with a large outburst in X-rays accompanied by increased emission at ultraviolet (UV), optical, near-infrared (nIR) and radio wavelengths \citep[for a review, see e.g.][]{mcclintock2006,belloni2010,fender2010,gallo2010,gilfanov2010}. 
These systems spend most of their time in a quiescent state, where the optical/nIR emission is emitted by the companion star or the cool accretion disk. 
The flux across the electromagnetic spectrum increases by orders of magnitude during outbursts, which are powered by increased accretion onto the black hole and can last weeks to months. 
The radio emission in these outbursts is ascribed to a relativistic jet, and sometimes these sources are named microquasars by analogy to radio-loud active galactic nuclei \citep{mirabel1992}.

During an outburst BHXBs go through several canonical X-ray spectral states, following a typical trajectory in the hardness-intensity diagram \citep[HID; e.g.][]{homan2001,homan2005b,belloni2010}. 
Typically, the source will first become significantly brighter while the spectrum is relatively hard and power-law dominated (hard state), after which the source will become spectrally soft and thermal dominated (soft state); and it eventually dims and evolves back to the hard and then quiescent states. 
The various spectral states can also be associated with different time variability behavior and the presence of certain types of quasi-periodic oscillations \citep[QPOs; e.g.][]{wijnands1999,casella2005,motta2011}. 
There are various classifications in the literature for the different states, based on spectral and timing behavior. 
In this paper we follow \citet{homan2005b} and \citet{belloni2010} in identifying the hard-intermediate state (HIMS) and soft-intermediate state (SIMS) in between the hard and soft states \citep[see][for an alternative classification]{mcclintock2006}. 

The X-ray emission is thought to be produced in the accretion disk and/or a hot corona above the disk, and the UV, optical and nIR emission is also produced in the accretion disk, either intrinsically \citep[e.g.][]{shakura1973} or by reprocessing of X-rays in that same region \citep[e.g.][]{vanparadijs1994}. 
At nIR to optical wavelengths there may also be a contribution from the jet that dominates the radio emission \citep[e.g.][]{corbel2002,russell2006}. 
The latter emission is usually described by a flat spectrum and it is bright during the initial hard state. 
In some models there is also a contribution from the jet at X-ray frequencies \citep[e.g.][]{markoff2004}. 
During the transition to the soft state the radio emission is quenched, and in some sources discrete ejecta are launched and their radio emission can be spatially resolved \citep[e.g.][]{fender2004}.

On 2010 September 25 a new transient was discovered with the Burst Alert Telescope (BAT) onboard the {\it Swift} satellite, which was initially identified as a gamma-ray burst (GRB) and named GRB\,100925A \citep{manganogcn11296}. 
The source was independently detected as a peculiar hard X-ray emitting source by the Gas Slit Camera (GSC) of the Monitor of All-sky X-ray Image (MAXI) instrument onboard the International Space Station, and was designated as \maxi \citep{negoroatel2873}. 
Soon after its discovery it was realized that this new transient was not a GRB \citep{kanngcn11299,xugcn11303}, and observations with the X-shooter spectrograph at the ESO Very Large Telescope (VLT) identified the source as an X-ray binary \citep{deugartepostigogcn11307,kaur2012}. 
The nature of the source was confirmed by the Rossi X-ray Timing Explorer ({\it RXTE}), classifying it as a black hole candidate \citep{kalamkaratel2881} based on the identification of its low-frequency QPOs. 
\maxi evolved through the X-ray hardness-intensity diagram in a manner characteristic of low-mass X-ray binaries, and the presence of type-B and type-C QPOs, and the way in which they evolve during the outburst, have provided evidence for the black hole nature of the compact object \citep{kalamkar2011,munozdarias2011}. 

Further observations at X-ray and soft gamma-ray energies with {\it Swift} \citep{kenneaatel2877}, {\it INTEGRAL} \citep{kuulkersatel2912}, {\it RXTE} \citep[e.g.][]{belloniatel2926}, MAXI and {\it XMM-Newton}, have shown that there are dips in the light curves recurring at a period of $2.414\pm0.005$ hours \citep{kuulkers2013}. 
These have been interpreted as absorption dips at the orbital period of the system, which makes \maxi the shortest period black hole candidate known to date \citep{kuulkers2012,kuulkers2013,kennea2011}. 
For absorption dips to occur, the inclination angle of the accretion disk with respect to the line of sight is relatively well constrained at $\sim65-80$\degr. 
This allows for an estimate of the binary orbital separation, and mass and radius of the donor star. 
The binary system appears to be compact (orbital separation of $>1.33~\rm{R}_{\odot}$) and the companion is suggested to be an M5 or M2 dwarf star \citep{kuulkers2013,kong2012}. 

After the source discovery we initiated a broadband follow-up campaign. 
Besides many of the aforementioned observations at X-ray and soft gamma-ray energies, we observed the source in various UV, optical and nIR bands, and at sub-millimeter and several radio frequencies. 
A varying optical source was detected with the {\it Swift} UltraViolet and Optical Telescope \citep[UVOT;][]{marshallgcn11298}, and in the R-band with BOOTES-2 and IAC80 \citep[][]{jelinekgcn11301}, while optical variability on minute timescales was found with the Faulkes Telescope \citep[][]{russellatel2884}. 
At sub-millimeter wavelengths \maxi was detected with the Atacama Pathfinder EXperiment \citep[APEX;][]{deugartepostigogcn11304}, 
and at radio frequencies with the Westerbork Synthesis Radio Telescope \citep[WSRT;][]{vanderhorstgcn11309}. 
We followed the evolution of the outburst during various transitions at X-ray \citep[][]{belloniatel2927,shaposhnikovatel2951,munozdariasatel2999} and radio \citep[][]{vanderhorstatel2918,paragiatel2906} wavelengths. 

Here we present the results of our observing campaign across the electromagnetic spectrum, focusing on the broadband modeling of several epochs for which we have well covered spectral energy distributions (SEDs) including observations at several radio frequencies, and on correlations between the different spectral regimes. 
In Section~\ref{sec:obs} we describe all our broadband observations and data analysis, 
and in Section~\ref{sec:broadlcs} we present the resulting light curves. 
Our broadband spectral modeling is detailed in Section~\ref{sec:modeling}, 
while we discuss the radio-X-ray and nIR/optical/UV-X-ray correlations in the context of other BHXBs in Section~\ref{sec:radioxraycor}. 
We summarize and conclude in Section~\ref{sec:conclusions}. 
All uncertainties in measured quantities and modeling parameters are quoted at the $1\sigma$ confidence level.

\section{Observations \& Data Analysis}\label{sec:obs}

\subsection{Radio}
\begin{table*}
\scriptsize
\begin{center}
\caption{Radio and sub-millimeter observations of \maxinos, with $\Delta$T the number of days after MJD 55464.0 (the source discovery date). 
WSRT$^*$ indicates WSRT observations taken as part of EVN e-VLBI runs \citep{paragi2013}. 
Upper limits are given at the $3\sigma$ level.}
\label{tab:radio}
\renewcommand{\arraystretch}{1.1}
\begin{tabular}{|l|c|c|c|c|c|c|c|c|c|c|} 
\hline
Epoch & $\Delta$T & Obser- & 0.61 GHz & 1.4 GHz & 2.3 GHz & 4.9 GHz & 8.5 GHz & 22 GHz & 43 GHz & 345 GHz \\
(MJD) & (days) & vatory & (mJy) & (mJy) & (mJy) & (mJy) & (mJy) & (mJy) & (mJy) & (mJy) \\
\hline\hline
55464.96 $-$ 55465.08 & 1.02 & APEX & ... & ... & ... & ... & ... & ... & ... & 15.8$\pm$3.0 \\
55465.59 $-$ 55465.82 & 1.70 & WSRT & ... & ... & ... & 5.39$\pm$0.05 & ... & ... & ... & ... \\
55467.05 $-$ 55467.13 & 3.09 & VLA & ... & ... & ... & 10.30$\pm$1.03 & ... & 10.00$\pm$1.05 & ... & ... \\
55467.17 $-$ 55467.34 & 3.25 & ATCA & ... & ... & ... & 9.76$\pm$0.06 & 11.29$\pm$0.06 & ... & ... & ... \\
55467.51 $-$ 55467.83 & 3.67 & WSRT & ... & 6.95$\pm$0.14 & 6.68$\pm$0.08 & ... & ... & ... & ... & ... \\
55468.05 $-$ 55468.13 & 4.09 & VLA & ... & ... & ... & 9.88$\pm$0.30 & 10.03$\pm$0.31 & 11.81$\pm$0.71 & 11.19$\pm$0.59 & ... \\
55469.63 $-$ 55469.76 & 5.69 & WSRT$^*$ & ... & ... & ... & 9.77$\pm$0.10 & ... & ... & ... & ... \\
55470.06 $-$ 55470.15 & 6.10 & VLA & ... & ... & ... & 10.29$\pm$0.32 & 9.74$\pm$0.30 & 8.84$\pm$0.49 & 4.84$\pm$0.35 & ... \\
55471.50 $-$ 55471.82 & 7.66 & WSRT & ... & 9.02$\pm$0.16 & 9.40$\pm$0.08 & ... & ... & ... & ... & ... \\
55471.98 $-$ 55472.06 & 8.02 & VLA & ... & ... & ... & 9.23$\pm$0.28 & 7.55$\pm$0.42 & 7.88$\pm$0.42 & 3.74$\pm$0.40 & ... \\
55473.58 $-$ 55473.75 & 9.66 & WSRT$^*$ & ... & ... & ... & 3.65$\pm$0.09 & ... & ... & ... & ... \\
55473.97 $-$ 55474.06 & 10.01 & APEX & ... & ... & ... & ... & ... & ... & ... & 10.5$\pm$3.2 \\
55475.95 $-$ 55476.07 & 12.01 & APEX & ... & ... & ... & ... & ... & ... & ... & $<$6.0 \\
55476.48 $-$ 55476.80 & 12.64 & WSRT & ... & 1.15$\pm$0.10 & 1.26$\pm$0.06 & ... & ... & ... & ... & ... \\
55476.96 $-$ 55477.04 & 13.00 & VLA & ... & ... & ... & 0.63$\pm$0.03 & 0.59$\pm$0.03 & 0.41$\pm$0.07 & $<$0.10 & ... \\
55479.90 $-$ 55480.05 & 15.97 & APEX & ... & ... & ... & ... & ... & ... & ... & $<$6.3 \\
55480.97 $-$ 55481.01 & 16.99 & VLA & ... & ... & ... & 0.59$\pm$0.03 & 0.65$\pm$0.06 & ... & ... & ... \\
55482.98 $-$ 55483.03 & 19.00 & VLA & ... & ... & ... & 0.95$\pm$0.04 & 0.88$\pm$0.04 & ... & ... & ... \\
55484.46 $-$ 55484.78 & 20.62 & WSRT & ... & 2.23$\pm$0.12 & ... & 2.03$\pm$0.06 & ... & ... & ... & ... \\
55488.37 $-$ 55488.49 & 24.43 & GMRT & $<$0.42 & ... & ... & ... & ... & ... & ... & ... \\
55488.45 $-$ 55488.77 & 24.61 & WSRT & ... & $<$0.27 & $<$0.23 & ... & ... & ... & ... & ... \\
55488.99 $-$ 55489.03 & 25.01 & VLA & ... & ... & ... & 0.23$\pm$0.03 & $<$0.069 & $<$0.25 & ... & ... \\
55494.31 $-$ 55494.52 & 30.42 & GMRT & $<$0.15 & ... & ... & ... & ... & ... & ... & ... \\
55499.90 $-$ 55499.94 & 35.92 & VLA & ... & ... & ... & $<$0.072 & $<$0.066 & ... & ... & ... \\
55554.64 $-$ 55554.69 & 90.66 & VLA & ... & ... & ... & 0.081$\pm$0.016 & ... & ... & ... & ... \\
\hline
\end{tabular}
\end{center}
\end{table*}

We observed \maxi at radio and sub-millimeter frequencies ranging from 610 MHz to 345 GHz, using the WSRT, Karl G. Jansky Very Large Array (VLA), Australia Telescope Compact Array (ATCA), and Giant Metrewave Radio Telescope (GMRT). 
The results of our radio campaign are given in Table \ref{tab:radio}, and the upper panel of Figure \ref{fig:broadlcs} shows the light curves for four of our radio observing frequencies to illustrate the broadband radio evolution of the outburst.

Very Long Baseline Interferometry (VLBI) observations with the European VLBI Network (EVN) and Very Long Baseline Array (VLBA) are presented in \citet{paragi2013}. 
Resulting fluxes of these observations are not included here, since the source is resolved in those, but we have included the WSRT synthesis-array fluxes from the two EVN observations.

\subsubsection{Very Large Array}

We used the VLA to observe \maxi over multiple epochs at 5, 8, 22 and 43 GHz, starting on 2010 September 28 (MJD 55467) and ending on 2010 December 24 (MJD 55554), as detailed in Table~\ref{tab:radio}. 
The VLA was in the DnC configuration (maximum baseline of 1.9~km) for the observations from 28 September to 2 October, and in the C configuration (maximum baseline of 3.4~km) from 7 October through the end of our observations. 
We used the new WIDAR correlator with 2 sub-bands and 64 channels per sub-band covering a bandwidth of 128 MHz (i.e., 256 MHz in total). 
We selected 3C286 for our absolute flux calibrator and used two nearby calibration sources, J1707$-$1415 and J1658$-$0739, to apply complex gain calibration. 
For polarization measurements, we used OQ208 as our leakage calibrator.

We performed all of our calibration and reduction in the Astronomical Image Processing System \citep[AIPS;][]{wells1985}, using standard procedures for radio frequency interference excision, spectral channel selection, and calibration. 
For estimating the systematic uncertainties in our measured flux densities, we went by the guidelines outlined in the VLA Calibrator Manual. 
We added in quadrature systematic uncertainties of 3\% (4.9 and 8.5~GHz) and 5\% (22 and 43~GHz) of the measured flux densities. 
As with any new instrument, we experienced some ``teething trouble'' with the VLA. 
Our first observing run produced little useful data. 
We only managed to get time on \maxi but had no time on 3C286 for flux calibration and only 1 scan per band on J1658$-$0739. 
We used measurements of J1658$-$0739 from our second observing run to bootstrap fluxes from the first run. 
We also missed 3C286 at 8.5~GHz during our observing run on 2010 October 11 (MJD 55480) and had to bootstrap the flux calibration using J1707$-$1415. 
These problems resulted in significantly higher uncertainties for these two observations. 

We determined the degree of polarization in our 5 and 8~GHz observations in the first two weeks of the outburst. 
We did not detect any significant polarization in our measurements, with 3$\sigma$ upper limits varying from 1\% to 5\% in the first three epochs, and tens of percent in the fourth epoch. 
At later epochs the polarization limits are less constraining due to the reduced brightness of the source.

\subsubsection{Westerbork Synthesis Radio Telescope}

We performed observations of \maxi with the WSRT at 1.4, 2.3 and 4.9~GHz, using the Multi Frequency Front Ends \citep{tan1991} in combination with the IVC+DZB back end in continuum mode, 
with a bandwidth of 8x20 MHz at all observing frequencies. 
Complex gain calibration was performed using the calibrator 3C286 for all observations. 
The observations were analyzed using the Multichannel Image Reconstruction Image Analysis and Display 
\citep[MIRIAD;][]{sault1995} software package, except for the WSRT data that were obtained during VLBI observations, 
which were analyzed with AIPS. 
The flux uncertainties were determined with the MIRIAD task imfit, to which the rms noise in the image around \maxi was added in quadrature.

With the initial discovery of the radio counterpart of \maxi we also reported a high degree of linear polarization in the source \citep{vanderhorstgcn11309}. 
However, careful re-analysis of this first epoch, and also analysis of the other WSRT epochs at 4.9 GHz, does not show any significant polarization, consistent with the non-detections of polarized emission with the VLA.

\subsubsection{Australia Telescope Compact Array}

\maxi was observed with the ATCA for 4.5~hours on 2010 September 28 (MJD 55467). 
We used the Compact Array Broadband Backend \citep[CABB;][]{wilson2011} to observe simultaneously in two separate frequency bands, centered at 5.5 and 9.0~GHz. 
Each frequency band comprised 2048 channels, each of width 1~MHz, for a total observing bandwidth of 2048~MHz. 
The array was in its H75 configuration, with five closely-spaced antennas within 82~m, and a more distant sixth antenna 4.4~km away.

Data were converted to fits format using the MIRIAD software package, and then read into the Common Astronomy Software Applications (CASA) package for data editing, calibration and imaging. 
We used PKS\,1934$-$638 as the primary calibrator, and J1733$-$1304 as the secondary calibrator. 
One of the central antennas was shadowed during observations of the primary calibrator, so had to be flagged. 
After editing out bad data and performing external gain calibration, we made a naturally-weighted image of the field containing \maxinos, using the multi-frequency synthesis algorithm in CASA to accurately deconvolve sources with non-zero spectral slopes. 
Owing to the large hole in the {\it uv}-coverage arising from the isolation of the sixth antenna, we only used data from the inner antennas during the imaging process, providing an angular resolution of 65 and 42~arcsec at 5.5 and 9.0~GHz, respectively. 
However, a comparison with higher-resolution VLA data shows that there were no bright confusing sources within these large beams, so the relatively poor angular resolution should not affect our photometry. 

A second observation was made on 2010 October 1 (MJD 55470), with a central frequency of 2.1~GHz. Owing to the RFI affecting the majority of the observing bandwidth, and to the short duration of the observation (1.5~hours), we were unable to use these data to place any accurate constraints on the brightness of the target source.

\subsubsection{Giant Metrewave Radio Telescope}

GMRT observed the field of \maxi on 2010 October 19 and 25 (MJD 55488 and 55494, respectively) at an observing frequency of 610 MHz. 
Radio sources 3C286 and J1626$-$298 were used as flux and phase calibrators, respectively. 
The flux calibrator was observed for about 15 minutes at the start of the target observations. 
The phase calibrator and the field of \maxi were then observed alternately for 6 and 45 minutes. 
The analysis of the observations was carried out using AIPS, and resulted in non-detections at this frequency.

\begin{figure*}
\begin{center}
\includegraphics[width=0.84\textwidth]{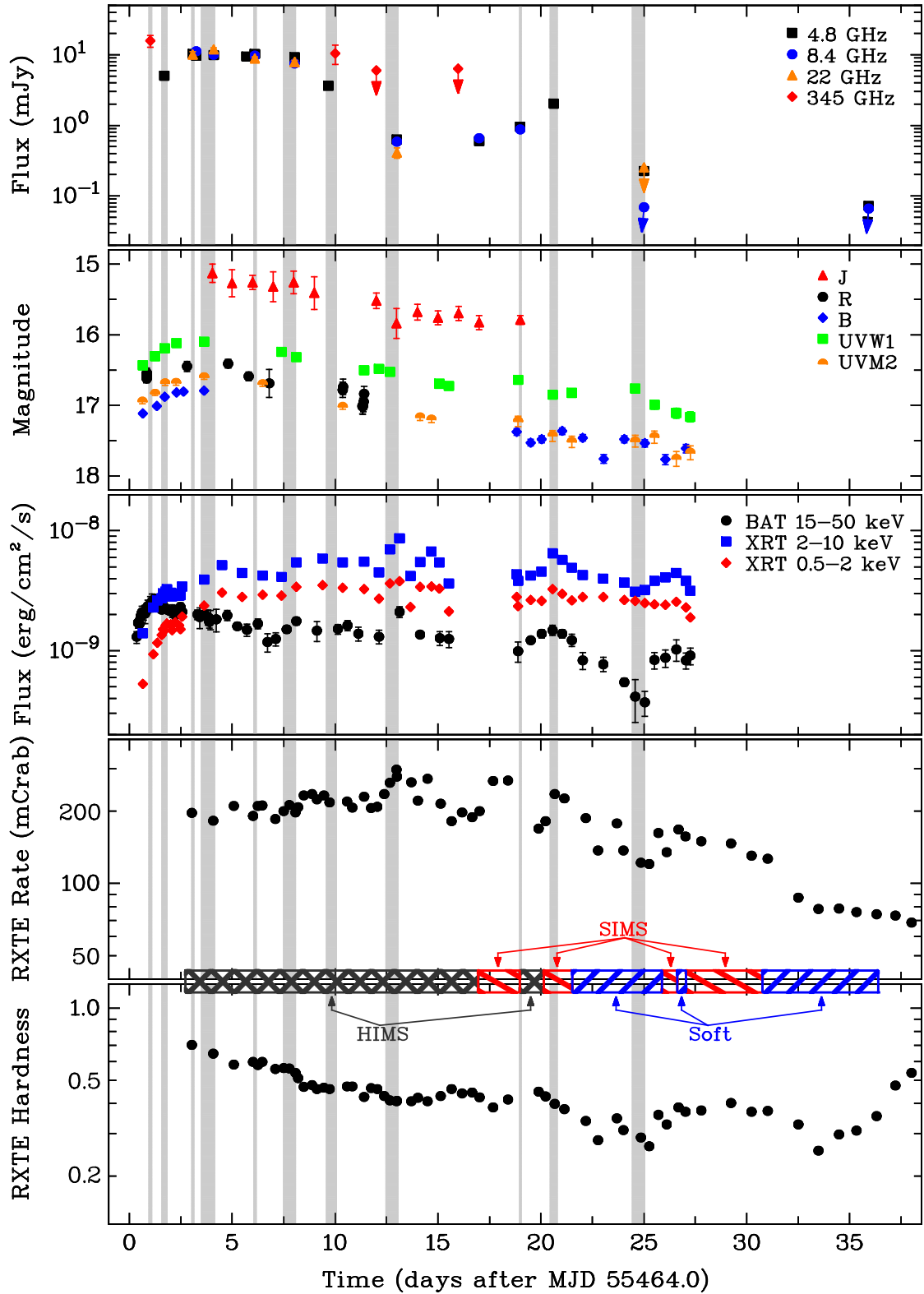}
\caption{Light curves at various radio and sub-millimeter wavelengths (top panel), in nIR/optical/UV bands (2nd panel from the top), from {\it Swift}/BAT and {\it Swift}/XRT (middle panel), and from {\it RXTE} ($2-15$~keV; 2nd panel from the bottom); and the {\it RXTE} hardness ratio ($6-15$~keV/$2-6$~keV; bottom panel). 
The vertical grey bars indicate the epochs for which we performed broadband SED modeling. 
The times when the source was in the various X-ray states (HIMS, SIMS or soft state) are also indicated \citep[following][]{munozdarias2011}. 
Note that the source was in the hard state in the first hours after the outburst onset, but the exact transition from hard state to HIMS is unknown due to the lack of {\it RXTE} observations. 
}
\label{fig:broadlcs}
\end{center}
\end{figure*}

\subsection{Sub-millimeter}

Continuum observations at 870 $\mu$m (i.e., 345~GHz) were carried out using the Large APEX BOlometer CAmera \citep[LABOCA;][]{siringo2009} installed on the Atacama Pathfinder EXperiment \citep[APEX;][]{lundgren2010}. 
Data were acquired at four epochs under good weather conditions (zenith opacity values ranged from 0.17 to 0.43 at 870 $\mu$m). 
Observations were performed using the wobbler on-off mode and data were reduced using the Bolometer Array (BoA) analysis software. 
The total on-source integration time of the four epochs was 8.6~hr. 
The telescope pointing was checked every hour, finding an rms pointing accuracy of 2.1~arcsec. 
Calibration was performed using observations of the primary calibrators Mars, Uranus and Neptune, as well as the secondary calibrators G10.62, IRAS16293 and G5.89. 
The absolute flux calibration uncertainty is estimated to be 11\%.

\subsection{Near-Infrared, Optical \& Ultraviolet}

The nIR (J- and H-band) observations were performed using the 1.3m~telescope 
(previously the 2MASS southern telescope) at the Cerro Tololo Inter-American Observatory (CTIO), 
which is currently operated by the Small \& Moderate Aperture Research Telescope System 
(SMARTS) consortium \citep[][]{subasavage2010}. 
We observed \maxi with an almost daily cadence between 2010 September 29 (MJD 55468) and October 14 (MJD 55483). 
The data were recorded by a Rockwell HgCdTe Astronomical Wide Area Infrared Imager. 
Multiple dithered frames were taken and then flat-fielded, 
sky subtracted, aligned, and average-combined using an in-house IRAF script. 
We used two nearby 2MASS stars in the field of \maxi as references. 
The average magnitudes of these comparison stars were used as a basis for differential photometry with respect to \maxinos. 
Additional late-time deep observations in the J-band were carried out with the 3.5m CAHA telescope on 24 July 2013 (MJD 56497) with a total integration time of 38 minutes, resulting in a significant detection of $21.05\pm0.17$~mag, $\sim3$~years after the outburst onset. 

Optical observations were carried out with the IAC80, 1.23m CAHA, 2.2m CAHA, 2.0m Liverpool, BOOTES-2/Telma and BOOTES-3/Yock-Allen telescopes. 
The IAC80 telescope is a 0.82m aperture facility, and the observations were done with the CAMELOT camera which is based on a 2k$\times$2k E2V CCD providing a pixel scale of 0.3~arcsec/pixel. 
The 1.23m CAHA, BOOTES-2/Telma and BOOTES-3/Yock-Allen facilities are Ritchey-Chretien telescopes. 
The 1.23m CAHA data were acquired with a 2048$\times$2048 SITE\#2b CCD which yields a pixel scale of 0.5~arcsec/pixel. 
Both BOOTES stations are equipped with identical cameras, an Andor iXon-889+ electron multiplying CCD of 1024$\times$1024 pixels, with a pixel scale 0.6~arcsec/pixel \citep[see][for more details on the BOOTES network of telescopes]{castrotirado1999}. 
The IAC80 and the 1.23m observations were carried out using standard Johnson/Cousins filters \citep{fukugita1995}, while the images of both BOOTES stations were acquired using Sloan filters \citep{fukugita1996}. 

The 2.0m Liverpool Telescope (LT) observations were made in the SDSS $i'$-band with RATCam, which yields a pixel scale of 0.135$^{\prime\prime}$/pix based on a 2048x2048 pixel EEV CCD42-40. 
Additional $g'r'i'z'$-band observations were carried out with the 2.2m Calar Alto telescope equipped with BUSCA. 
BUSCA is a multichannel camera which allows simultaneous direct imaging in four optical bands using three CCD485 Lockheed Martin plus one CCD485 backside thinned CCD (in the uv channel). 
The  BUSCA observations were done in 2$\times$2 binning and 1k$\times$1k windowing mode, yielding a pixel scale of 0.35~$\arcsec$/pix and a field of view of 6$\arcmin\times6\arcmin$.

The reduction of all the optical data was performed using standard procedures implemented in IRAF. 
Calibration of the Sloan data was carried out transforming the Johnson magnitudes of 22 field stars based on \citet{jordi2006}. 
The source looks relatively isolated in optical images: there are no entries in either the 2MASS \citep{skrutskie1995} or USNO-B1.0 \citep{monet2003} catalogues consistent with the position of \maxinos. 
Furthermore, observations with the Canada-France-Hawaii Telescope at $\sim18$ months after the start of the X-ray outburst result in a quiescent $r$ band magnitude of $\sim23.6$ \citep{kong2012}. 
We hence assume that the flux at the time of our analysis was dominated by the transient emission of \maxinos. 

Data from the {\it Swift} UV/Optical Telescope were downloaded from the HEASARC archive and pre-processed at the {\it Swift} Data Center \citep[see][]{breeveld2010}, 
and required only minimum user processing with the {\it Swift} {\tt FTOOLS} as follows. 
The image data from each filter, from each observation sequence, were summed using {\tt uvotimsum}. 
Photometry of the source in individual sequences was derived via {\tt uvotmaghist}, 
using an extraction region of radius 5\arcsec. 
{\tt XSPEC} compatible spectral files were created with that same region using {\tt uvot2pha}.
The obtained magnitudes are based on the UVOT photometric system \citep{poole2008}. 

In Table~\ref{tab:optical} we give the results of our ground-based optical follow-up campaign and the {\it Swift}/UVOT observations.

\subsection{X-rays}

The outburst of \maxi was observed by several X-ray satellites and instruments. 
In this paper we focus on the {\it Swift}, {\it RXTE} and MAXI observations. 

The light curves and spectra of the {\it Swift} X-Ray Telescope (XRT) were obtained from the XRT online tool, which offers science grade products \citep{evans2009}. 
{\it Swift} Burst Alert Telescope (BAT) data were downloaded from the HEASARC archive and initially processed with the {\tt FTOOL} {\tt batsurvey}, which applies standard corrections. 
8-channel spectra and response files were extracted and the standard spectral systematic error correction applied with {\tt batphasyserr}. 
All X-ray spectra were binned to have $\geq20$ counts per bin (with {\tt grppha}) so that errors would be approximately Gaussian and hence $\chi^{2}$ statistics would be valid.

We analyzed all 65 observations of \maxi in the {\it RXTE} archive. 
For each observation we extracted a background and dead-time corrected energy spectrum from the PCA standard mode, using only data from Proportional Counter Unit (PCU) 2 of the PCA, as it is the best calibrated and the only one which is always active. 
We used the standard HEASOFT {\it RXTE} software to create energy spectra. 
From these spectra, we extracted background-corrected count rates in the PCA channel bands $A=0-35$ ($2-15$~keV; total band), $B=0-13$ ($2-6$~keV; soft band) and $C=14-35$ ($6-15$~keV; hard band). 
The {\it RXTE} rate $A$, converted to Crab units for PCU~2 (conversion factor of 1~Crab~$=$~2284~counts/second), was used as {\it RXTE} rate in Figures~\ref{fig:broadlcs} and \ref{fig:hid}, while the hardness was defined as the ratio $C/B$. 

The daily count rate ($2-20$~keV) and hardness ($4-10$~keV~/~$2-4$~keV) as measured by the MAXI instrument have been taken from the MAXI website\footnote{http://maxi.riken.jp}.

\section{Broadband Light Curves}\label{sec:broadlcs}

Figure~\ref{fig:broadlcs} shows a broadband overview of the light curves that we obtained in our follow-up campaign of \maxinos. 
In this figure we show the light curves at various radio frequencies, nIR, optical and UV bands, and X-ray and soft gamma-ray energies, from the observations presented in Section~\ref{sec:obs}. 
We do not show light curves in all observing bands to avoid cluttering, but these selected observing bands illustrate the broadband evolution of the \maxi outburst well. 
For comparison we also show the {\it RXTE} count rate light curve and hardness evolution, and indicate the times when the source is in the various X-ray states (HIMS, SIMS or soft state), following the identification of these spectral states for \maxi by \citet{munozdarias2011}. 
We note that the source was only in the hard state during the first hours after outburst onset, before the start of the {\it RXTE} observations. 
Because there are no {\it RXTE} observations during the first two days, we can not identify the exact time of the transition between the hard state and the HIMS. 

It can be seen in Figure~\ref{fig:broadlcs} that the source evolved in different ways in the various parts of the electromagnetic spectrum. 
In almost all wavebands we witnessed the rise of the source brightness, except for our APEX sub-millimeter observations which do not have the required temporal sampling, and the nIR observations which started around the peak of the outburst at those wavelengths. 
The observed trends in the optical, UV and X-ray light curves have been described in \citet{kennea2011}, with a correlated fast rise and slow decay at these frequencies, while the late-time X-ray behavior has been discussed in \citet{homan2013}. 
At radio frequencies the source was already fairly bright at 1.7~days after the start of the outburst (MJD 55464.0), and reached maximum brightness around day~3. 
During the following 5 days the radio brightness was constant, after which there was a sudden large decrease of the radio emission \citep{vanderhorstatel2918}. 
This radio flux drop was followed several days later by a transition from the hard-intermediate to the soft-intermediate state at X-ray energies \citep{belloniatel2927}, and the radio flux started rising again after that. 
Other BHXBs have shown that at the transition to the SIMS, fast radio jets in the form of optically thin, discrete ejecta may emerge. 
This has, however, not been seen for \maxi \citep{paragi2013}. 
After the second rise, the flux dropped significantly again at day~20 after the beginning of the outburst, and subsequently it was only detected at very low flux levels \citep{millerjonesatel3358,jonker2012}.

\section{Broadband Spectral Modeling}\label{sec:modeling}

Given our rich broadband data set we performed broadband SED modeling. 
The grey vertical bands in Figure~\ref{fig:broadlcs} indicate the 11 epochs for which this was done. 
We made our selection of the epochs based on the times for which we had radio data available. 
In the X-ray bands we used {\it Swift}/XRT and {\it Swift}/BAT data. 

The broadband radio-to-X-ray spectra were fit within {\tt XSPEC (12.7.1)}, using $\chi^{2}$ statistics and accounting for interstellar extinction and absorption, which were modeled by {\tt redden} and {\tt tbabs}, respectively. 
The observed radio and optical flux densities $F_{\nu}$ at frequency $\nu$, were converted to flux per filter, $F_{\rm{filter}}$ in units of photons\,cm$^{-2}$\,s$^{-1}$. 
This was done via $F_{\rm{filter}} = 1509.18896 F_{\nu}$ $( \Delta\lambda/\lambda )$, where $\lambda$ and $\Delta\lambda$ are the effective wavelength and width of respective filters. 
This flux was then used to produce {\tt XSPEC} compatible files for spectral fitting, using the {\tt FTOOL} {\tt flx2xsp}.

At all epochs the data were well fit by a phenomenological broken power law at radio frequencies, and by a physically motivated irradiated disk model ({\tt diskir}) at optical/UV and X-ray frequencies \citep{malzac2005,gierlinski2008,gierlinski2009}. 
In the latter model the X-ray emission consists of thermal emission from the accretion disk and a hard tail caused by Comptonisation of soft seed photons in a hot plasma of energetic electrons. 
The Comptonised emission in turn illuminates the disk, of which a large fraction gets reflected, but also a significant fraction of the photons are reprocessed and add to the disk emission. 
The parameters of this model are: the unilluminated disk temperature $kT_{disk}$; the power-law photon index $\Gamma$ of the Compton tail; the ratio of luminosity in the Compton tail to that in the unilluminated disk, $L_{C}/L_{disk}$; the fraction of bolometric flux which is thermalised in the outer disk, $F_{out}$; the log of the ratio of outer to inner disk radius, $r_{out}$; and a normalisation, $Norm$, dependent on the apparent inner disk radius. 
We note that it has been shown that spectra can be fit by an irradiated disk model even in the hard state \citep[e.g.][]{miller2006,rykoff2007}.

The $\chi^{2}_{\nu}$ of the fits presented in the next section ranges from 1.1 to 2.0 for $\sim 10^{3}$ degrees of freedom, which are relatively high but given the disparity of the data (combination of many different instruments, epochs and bands) we deemed those acceptable. 
Furthermore, any variations from the model show no underlying structure, only noise-like variations.

\begin{figure}
\begin{center}
\includegraphics[viewport=0 220 554 794,clip,width=\columnwidth]{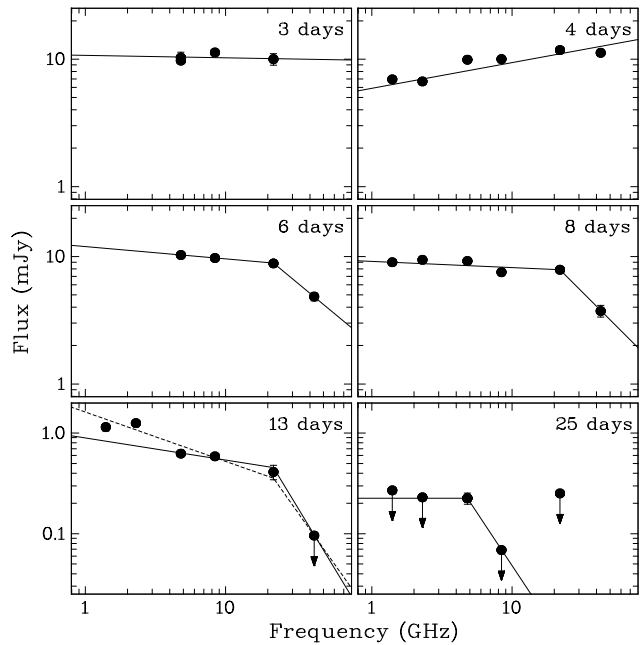}
\caption{Radio spectra of \maxi for the 6 epochs for which we have at least 3 observing bands. 
The lines indicate the (broken) power-law fits to the radio spectra.  The solid and dashed lines for the spectrum at day~13 show the fits with and without including the low-frequency WSRT data below 3~GHz.}
\label{fig:radioseds}
\end{center}
\end{figure}

\begin{figure}
\begin{center}
\includegraphics[angle=0,width=\columnwidth]{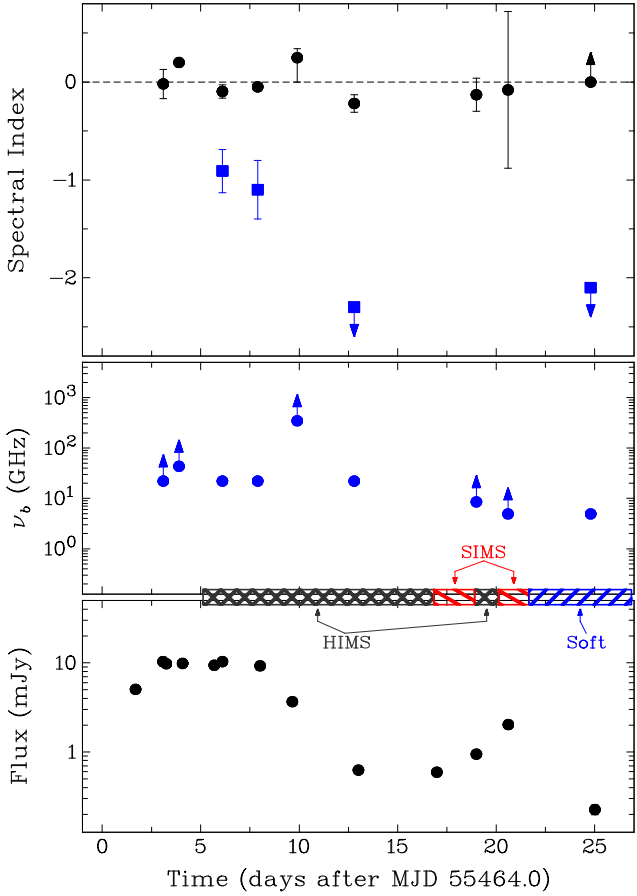}
\caption{Evolution of parameters from the broken power-law fits to the radio SEDs indicated in Figure~\ref{fig:broadlcs} except for the first two epochs. 
The top panel shows the spectral index, with the black circles indicating the low frequency power-law slopes, and the blue squares the high frequencies in those epochs that are best described by a broken power law. 
The middle panel shows the evolution of the spectral break for all epochs, including those for which we can only put upper or lower limits. 
In the lower panel the $4.9$~GHz light curve and the X-ray spectral states are displayed for comparison purposes.}
\label{fig:radiospecindex}
\end{center}
\end{figure}

\subsection{Modeling Results}\label{sec:modresults}

In our modeling the radio emission and the nIR to X-ray emission appear to be due to separate components. 
Figure~\ref{fig:radioseds} shows the radio spectra for 6 of our 11 epochs for which we have at least 3 radio bands. 
The behavior of the radio SEDs seems more complicated than a single flat or slightly-inverted spectrum across all radio bands (i.e. a spectral index $\sim~0$). 
We performed single and/or broken power-law fits to the radio SEDs for the epochs indicated in Figure~\ref{fig:broadlcs} except for the first two epochs in which there was only one observing frequency. 
From Figure~\ref{fig:radioseds} it can be seen that there are epochs where a single power law is not sufficient to describe the spectrum, which is the case for 4 out of the 9 epochs we fit. 
For the latter epochs we show both spectral indices in Figure~\ref{fig:radiospecindex}, while for the other 5 epochs we only show a single spectral index. 
In the broken power law fits we fixed the break at 22~GHz for the SEDs at 6, 8 and 13~days after MJD 55464.0, while at 25~days it was fixed at 4.9~GHz. 
In the former three epochs these break frequencies were fixed because the spectrum is consistent with a single power law up to 22~GHz and there is only one data point at 43 GHz to constrain the break, which makes the parameters of a free broken power-law fit hard to constrain. 
In fact, the break could be at any frequency in between 22 and 43~GHz and thus the high-frequency spectral indices can be steeper that the ones plotted in Figure~\ref{fig:radiospecindex}. 
At day~25 there is only one detection at 4.9~GHz, hence the lower limit on the low-frequency spectral index and the upper limit on the high-frequency spectral index. 
We would like to point out, however, that the observations are not strictly simultaneous, so the 4.9~GHz detection could be a short flare with a timescale of hours. 
For those epochs where a break is not required at radio frequencies, we show lower limits on the break frequency in Figure~\ref{fig:radiospecindex}. 
In our broadband radio-to-X-ray fits we placed those breaks arbitrarily at $10^{-5}$~keV ($\approx2.4\times10^{12}$Hz), with a steep cut-off.

\begin{figure}
\begin{center}
\includegraphics[angle=0,width=\columnwidth]{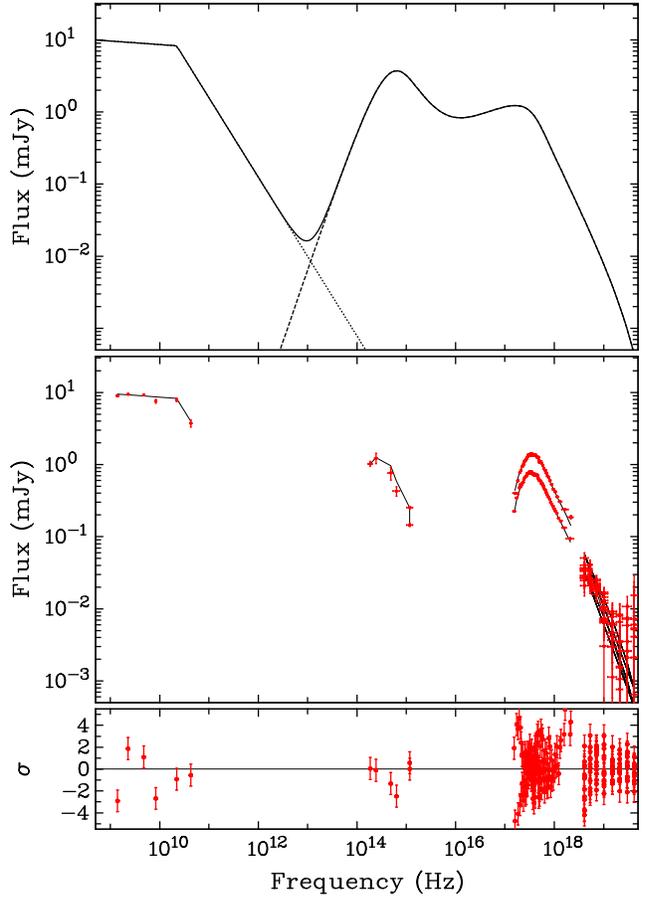}
\caption{Example model plots for the broadband SED at day~8 after the outburst onset. 
The top panel shows the spectrum of the best fit model for this particular epoch, with the dotted line showing the broken power-law fit to the radio data, the dashed line the irradiated disk model fit, and the solid line the total spectrum. 
The middle panel shows the fit of the unfolded spectrum, with the data in red symbols and the best fit model in black lines. 
In the bottom panel we show the deviations from the best fit model.}
\label{fig:fitepoch10} 
\end{center}
\end{figure}

An example fit to one of the epochs and the broadband model used here are shown in Figure~\ref{fig:fitepoch10}. 
$E(B-V)$ displays no significant variability with an average of $0.595 \pm 0.008$, consistent with the Galactic value of $E(B-V) = 0.599$ \citep{schlegel1998}. 
This is significantly larger than the value of $E(B-V)\simeq0.35$ derived by \citet{kaur2012}. 
The measured values of interstellar absorption are consistent with being constant with an average of $N_{{\rm H}}=(0.319\pm0.009)\times10^{22}$\,cm$^{-2}$. 
This value is greater than the expected Galactic absorption of $N_{{\rm H}}=0.17\times10^{22}$\,cm$^{-2}$ \citep{kalberla2005} but less than that obtained by \citet{kennea2011}. 
It should be noted, however, that \citet{kennea2011} fit a different model and only to the X-ray data, while we include nIR/optical/UV data which add additional constraints to the thermal emission. 

\begin{figure}
\begin{center}
\includegraphics[angle=-90,width=\columnwidth]{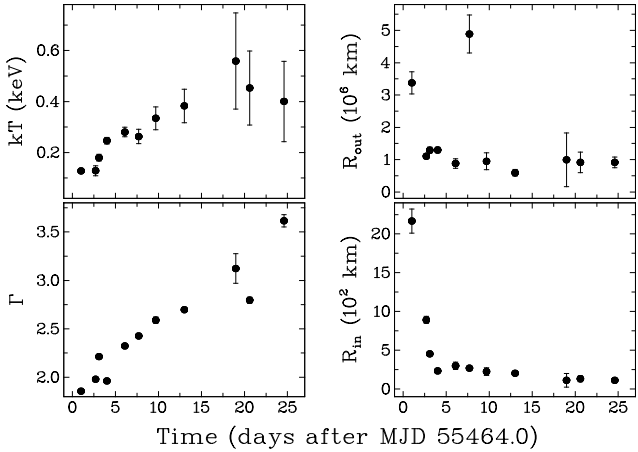}
\caption{Evolution of the disk temperature (top left), photon index of the Compton tail (bottom left), outer disk radius (top right), and inner disk radius (bottom right). These parameters are based on spectral fits to the broadband SEDs indicated in Figure~\ref{fig:broadlcs}, using the irradiated disk model.}
\label{fig:optxrayfits} 
\end{center}
\end{figure}

Figure~\ref{fig:optxrayfits} shows the evolution of the well-constrained parameters of the irradiated disk model. 
Although the fit parameter values we obtain are slightly different than the ones from \citet{kennea2011} and \citet{yamaoka2012}, the trends in their evolution are similar. 
The disk temperature, $kT_{disk}$, rises from an initial temperature of $0.13$\,keV to a relatively stable value of $\sim0.3$\,keV, while the photon index of the tail rises from an initial value of $\sim2.0$ to a peak of $\sim3.6$. 
As the disk temperature rises, the contribution that the Compton tail makes to the disk luminosity, $L_{C}/L_{disk}$, decreases, although this is poorly constrained on many epochs, and at late times is consistent with zero. 
Likewise, the fraction of bolometric flux which is thermalised in the outer disk, $F_{out}$, decreases although this is also poorly constrained. 
Since $L_{C}/L_{disk}$ and $F_{out}$ are poorly constrained, we have not plotted them in Figure~\ref{fig:optxrayfits}. 
The ratio of outer to inner disk radius, $r_{out}$, increases with time which is due to the fact that the inner disk radius is decreasing. 
Instead of displaying $r_{out}$ and $Norm$, we show the physical inner radius $R_{in}$ and outer radius $R_{out}$ in Figure~\ref{fig:optxrayfits}. 
The physical radius $R_{in}$ is related to the apparent radius $r_{in}$ of our fit by $R_{in}\approx1.19 r_{in}$ \citep[see, e.g.][]{soria2007}. 
We have determined $r_{in}$ at all the epochs from the fit values of $Norm$, assuming a source distance of 6\,kpc (see Section~\ref{sec:radioxraycor}) and an inclination of 70\arcdeg. $R_{out}$ was derived from $R_{in}$ and the ratio $r_{out}$. 
In the next section, we will discuss further the trends in the fitted parameters.

\subsection{Implications of Spectral Modeling}

\subsubsection{Radio SEDs \& Spectral Breaks}

\begin{figure}
\begin{center}
\includegraphics[angle=-90,width=\columnwidth]{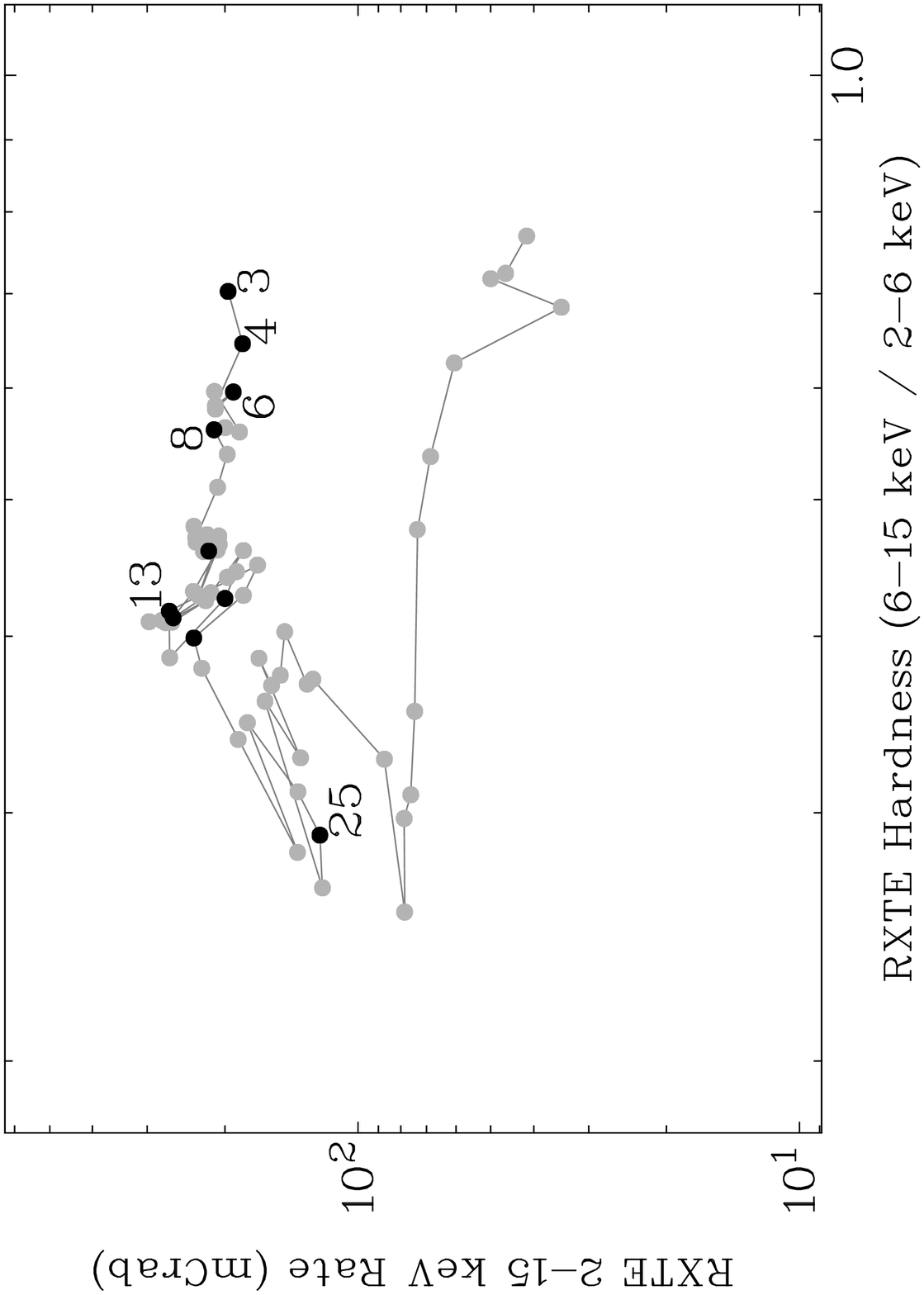}\vspace{0.3cm}
\includegraphics[angle=-90,width=\columnwidth]{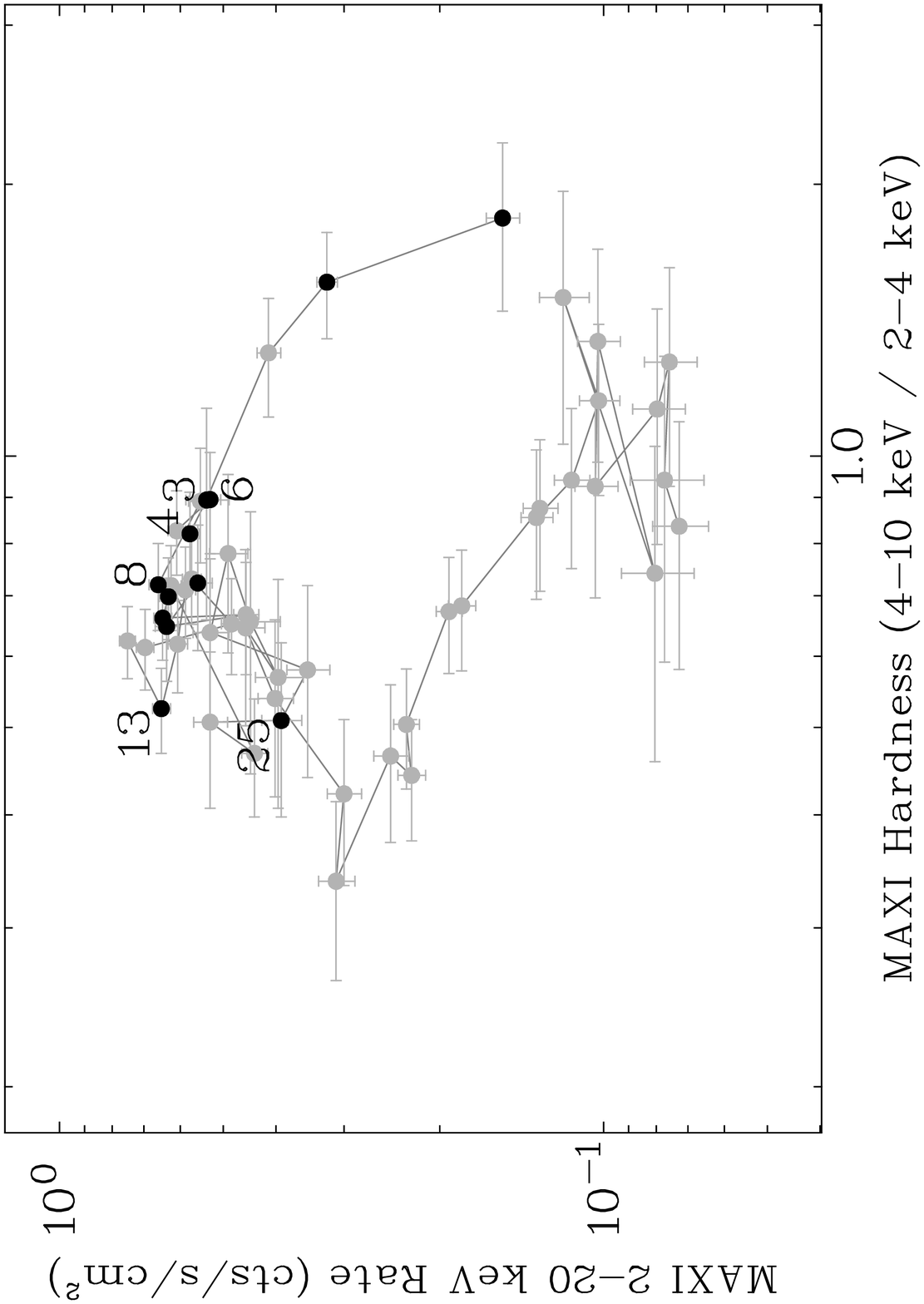}
\caption{Hardness-intensity diagram for the RXTE (top) and MAXI (bottom; up to MJD 55515) data. 
The black dots indicate the epochs of X-ray observations for which we have (quasi-) simultaneous radio data. 
We have indicated the epochs on days 3, 4, 6, 8, 13 and 25 from the radio spectra shown in Figure~\ref{fig:radioseds}. 
Note that RXTE was not observing \maxi from the outburst onset, hence there is one less black dot in the top panel compared to the bottom panel.}
\label{fig:hid}
\end{center}
\end{figure}

In this paper we present several epochs with broadband radio SEDs. 
To help place our SED epochs in the overall context of the behavior of the outburst, we indicate them on the HIDs of both the {\it RXTE} and MAXI data (Figure~\ref{fig:hid}). 
In this figure we have indicated the X-ray observation epochs for which we performed broadband SED modeling with (quasi-) simultaneous radio data. 
\maxi was already in the HIMS when the RXTE observations started, but from the MAXI HID we can deduce that the source was transitioning from the hard state to the HIMS during the first radio observations, $1-2$~days after the outburst onset. 
Our first radio SED with good broadband coverage is at $\sim3$ days after the onset of the X-ray outburst. 
The radio spectral index $\alpha$, where $F_{\nu}\propto\nu^{\alpha}$, is flat or inverted during the first 10 days over more than an order of magnitude in observing frequency, namely from 1.4 to 22~GHz, as one would expect from a synchrotron emitting, partially self-absorbed jet \citep[][]{fender2001} and in full agreement with the observed VLBI structure \citep[see][and the references therein]{paragi2013}. 
The spectrum is not constant over that time span though, with an inverted spectrum at day~4 with $\alpha=0.20\pm0.02$, while it is flat at other epochs (see Figure~\ref{fig:radiospecindex}). 
More striking is the sharp spectral break between 22 and 43~GHz that we find on days~6 and 8. 
After $\sim$10 days the radio flux drops significantly, but the spectrum is inverted even up to the sub-millimeter bands. 

At day~13 the spectrum appears more complicated, with a sharp spectral break above 22~GHz, and a best-fitting spectral index of $\alpha=-0.49\pm0.07$ at lower frequencies.  
We note, however, that the 1.4 and 2.3 GHz WSRT data were taken roughly 8 hours prior to the higher-frequency VLA data used to construct the spectrum shown in Figure~\ref{fig:radioseds}, and since the source was undergoing a rapid decay, may have overestimated the true low-frequency emission at the time of the VLA observations.  
Furthermore, any variations from a compact, partially self-absorbed jet will be detected later at lower frequencies, since the lower-frequency emission comes from further downstream in the jets.  
Thus, for a rapidly-decaying compact jet, we might expect the lowest-frequency emission to be slightly overestimated, relative to the power law seen at higher frequencies.  
Combined, these two effects could explain the deviation of the low-frequency spectrum on day $\sim13$ from a pure power law.  
To mitigate against this effect, we refitted the radio spectrum using only the VLA data between 4.9 and 22~GHz, to give a low-frequency spectral index of $\alpha=-0.22\pm 0.09$, with a break to $\alpha<-2.3$ above 22~GHz.  

When the radio flux increases again and the source moves into the SIMS, the spectrum remains flat, but we do not have the spectral coverage at high radio frequencies to check for a spectral break. 
The radio measurements at day~25 are also consistent with being flat or inverted, but then with a sharp break in between 4.9 and 8.5~GHz. 

From the radio light curves in Figure~\ref{fig:broadlcs} and the spectral index evolution in Figure~\ref{fig:radiospecindex} it appears that the decrease in the radio emission around day~13 after the outburst onset did not coincide with a significant steepening of the spectrum, as observed in other BHXBs \citep[e.g.][]{kuulkers1999,fender2004,millerjones2012}.  
The low-frequency radio emission remained consistent with a partially self-absorbed compact jet throughout the HIMS, with no evidence for ejection events, in agreement with the absence of ejecta in the VLBI images of \citet{paragi2013}.

At higher frequencies, however, the radio spectrum exhibited a high-frequency break above 22~GHz from day $\sim6$ until at least day $\sim13$, and the optically thin spectrum appeared to steepen with time. 
This spectral break can be interpreted as the transition from optically thick and partially self-absorbed to optically thin synchrotron emission, and has previously been observed in several other BHXBs in hard X-ray states although usually at much higher frequencies \citep{russell2013}. 
Recently, \citet{russell2013b} have shown that the BHXB MAXI\,J1836$-$194 has an inverted spectrum with a break towards a steep spectrum moving from the mid-infrared to lower frequencies as the source softens in the X-ray regime, and then back to higher frequencies as it gets harder again. 
With sufficiently dense monitoring over a broad radio frequency range one could be able to see the spectral break pass through the radio bands, which may explain what is happening between days~6 and 13 in \maxi \citep[i.e. the inverse of what was seen during the reverse transition in GX339-4 by][]{corbel2013a}. 
There appears to be a relatively rapid increase in the break frequency on day~10, shifting up to $>345$~GHz briefly before dropping back to 22~GHz on day 13. 
Since this is based on a spectrum consisting of flux density measurements at two frequencies, 4.9 and 345~GHz, this break frequency limit is quite uncertain. 
An inverted spectrum would shift it to a lower value, e.g. with a spectral index of 0.5 the break is at $\sim170$~GHz, 
lower than the limit of 345~GHz but still significantly larger than 22~GHz. 
Nonetheless, the evolution of the break frequency at that time is fast, but this is not unprecedented: in GX\,339-4 the spectral break shifted in the mid-infrared by more than an order of magnitude on timescales of hours when it was in the hard state \citep{gandhi2011}. 
Given that the radio-emitting part of the jet is significantly larger than the infrared-emitting region, strong variability on a timescale of days in the case of \maxi could be expected. 
Alternatively, this could be a short-lived flare at sub-millimeter frequencies. 
Flares get smoothed out and have reduced amplitude at lower frequencies, which could explain the lack of a flare at 4.9~GHz. 

The radio light curve shows an increase in flux between day~17 and 21, while the spectrum remains flat. 
In Figures~\ref{fig:broadlcs} and \ref{fig:radiospecindex} it can been seen that this coincides with a state transition (back and forth between the SIMS and the HIMS), consistent with what has been found for other BHXBs \citep[e.g.][]{corbel2013a,corbel2013b,russell2013b}. 
From these observations it seems that the jet started quenching in the HIMS and finished in the SIMS, but then recovered through a second HIMS and peaked in the second SIMS phase, before quenching again after a transition into the soft state. 
These changes in the radio emission coming from the jet and the X-ray state changes suggest a strong link between the accretion flow and the jet. 
It also shows that reaching the SIMS is no guarantee of a full transition to a jet ejection event, because there is no major radio flare when the sources moves into the soft state, and no discrete ejecta were detected in the VLBI observations \citep[][]{paragi2013}. 
Some other BHXBs in outburst also did not have ejection events, for instance Cyg~X$-$1 \citep[][]{rushton2012}, but that source did not trace out a canonical track in the HID, while \maxi did \citep{munozdarias2011}. 
The soft state in \maxinos, however, was slightly harder than seen in BHXBs with ejection events, and the minimum fractional variability in the soft state was slightly higher \citep[a minimum of 3\%, and rising to 8\% when the X-ray emission was softest;][]{munozdarias2011}. 
The authors attributed this to the high inclination of the system, but if it was instead intrinsic, the implied difference in the behaviour of the accretion flow could be related to the lack of jet ejection events. 
\citet{millerjones2012} have shown, however, that the ejection event occurs well before reaching the soft state, at the HIMS/SIMS transition \citep[see also][]{fender2009}. 
Since the behaviour of \maxi was relatively standard in the HIMS and SIMS, it seems unlikely that the unusual behaviour in the soft state could be causally related to the lack of ejection events.

\subsubsection{Physical Constraints from the Irradiated Disk Model}

In Figure~\ref{fig:optxrayfits} we show the evolution of the parameters of the irradiated disk model, in particular the disk temperature, the photon index of the Comptonised tail, and the inner and outer radius of the accretion disk. 
From this figure we can see that the source is getting spectrally softer (steeper photon index), the disk temperature is increasing, and the inner radius is decreasing, while the outer radius is fairly constant. 
The latter is not true for the first and sixth epoch, where we have only few nIR/optical data points to constrain the outer radius reliably. 
Excluding those epochs, the outer radius has a value of $\approx0.9-1.6\times10^{6}$~km, physically realistic considering the orbital separation \citep{kuulkers2013} and consistent with the position of the Lagrangian point $>10^{6}$~km. 
The inner radius of the disk is decreasing in time from $\sim890$ to $\sim110$~km, as one would expect in the standard picture of the disk evolution during a BHXB outburst. 
From the final value of $R_{in}$ (at day~25 after the outburst onset) we can put an upper limit on the black hole mass by assuming that $R_{in}$ is larger than the innermost stable circular orbit, $R_{ISCO}\equiv6GM/c^{2}$~km for a non-rotating, Schwarzschild black hole. 
The resulting black hole mass of \maxi is $<12M_{\sun}$, consistent with the mass derived by \citet{yamaoka2012}. We note that these mass estimates would be larger if the black hole were spinning and they should be treated with caution since they are model-dependent.

\section{Radio-X-ray \& Optical-X-ray Correlations}\label{sec:radioxraycor}

\subsection{Radio versus X-rays}

\begin{figure}
\begin{center}
\includegraphics[angle=-90,width=\columnwidth]{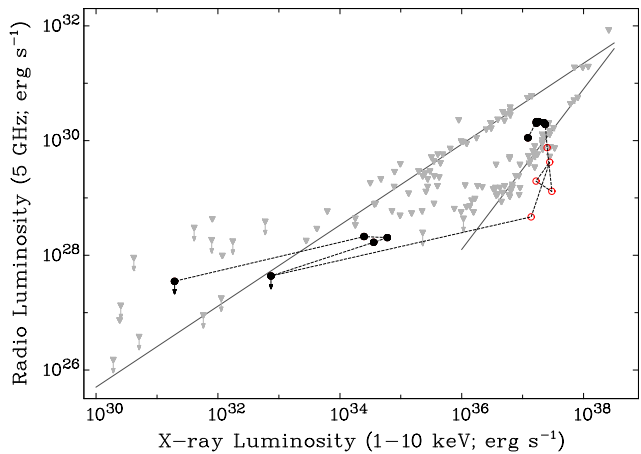}
\caption{Correlation between the radio and X-ray luminosity during the outburst of \maxinos, compared to other BHXBs (grey symbols). 
The grey lines indicate the two tracks in the radio-X-ray correlation with slopes of $\sim0.7$ and $\sim1.4$. 
The \maxi luminosities are calculated for a source distance of 6~kpc. 
The black solid circles indicate the early times of the outburst when the radio emission in \maxi had not dropped significantly yet (up to day~8 after the outburst onset) and the late times when the source is almost in quiescence ($>135$~days), while the red open circles indicate the times in between.}
\label{fig:radioxraycor}
\end{center}
\end{figure}

In Section~\ref{sec:modresults} we have discussed the related behavior between the radio and X-ray regime for \maxinos, and the implied link between the accretion flow and the jet. 
We explore this further by investigating the correlation between the radio and X-ray luminosities throughout the outburst. 
This correlation was originally suggested for GX\,339$-$4 when it was in the hard state \citep{hannikainen1998} and described by a power law with index $\sim0.7$ extending 3 orders of magnitude in luminosity with a turnover at high X-ray luminosities caused by a quenching of the radio emission in the soft state \citep{corbel2003,gallo2003}. 
This correlation has not only been found and studied in detail for individual sources \citep[e.g.][]{corbel2008,corbel2013a}, but also for the sample of BHXBs as a whole \citep[e.g.][]{gallo2003}. 
In recent years a more complex picture has revealed itself, with two distinct tracks in the radio/X-ray luminosity plane \citep[e.g.][]{coriat2011,gallo2012}. 
The second track lies below the first one and has a steeper power-law index of $\sim1.4$. 
For a few sources it has been shown that there is a transition from the lower to the upper track at low luminosities \citep{coriat2011,ratti2012}. 

Figure~\ref{fig:radioxraycor} shows the correlation for \maxinos, and a comparison with other BHXBs in the hard state. 
Besides the radio observations presented in this paper, we included the late-time data from \citet{jonker2012}. 
The latter paper focused mainly on the late-time evolution when the source was close to quiescence and also showed the early radio fluxes that had been reported in the literature \citep{vanderhorstatel2874,paragiatel2906}. 
To calculate the luminosities we adopted a distance to \maxi of 6~kpc, since the distances estimated in various ways span a range from 4 to 8~kpc \citep{kennea2011,millerjonesatel3358,kaur2012,kuulkers2013}. 

As shown by \citet{jonker2012}, \maxi was on the lower track early in the outburst, and once the source had reached quiescent levels, it seemed to transition to the upper track, although we note that the lowest-luminosity points have only radio upper limits. 
With our extensive data set during the early phases of the outburst we show in Figure~\ref{fig:radioxraycor} that the source gradually evolved off the correlation during the HIMS over the course of several days, as seen before in GX 339-4 \citep[][]{corbel2013b} and expected for sources which are not in the hard state. 
However, the motion in the radio/X-ray plane was not monotonic during this radio quenching, with the radio emission fading and recovering between days 6 and 20, due in part to the repeated transitions between the HIMS and the SIMS.

\subsection{Optical versus X-rays}

\begin{figure}
\begin{center}
\includegraphics[viewport=0 0 460 783,clip,width=\columnwidth]{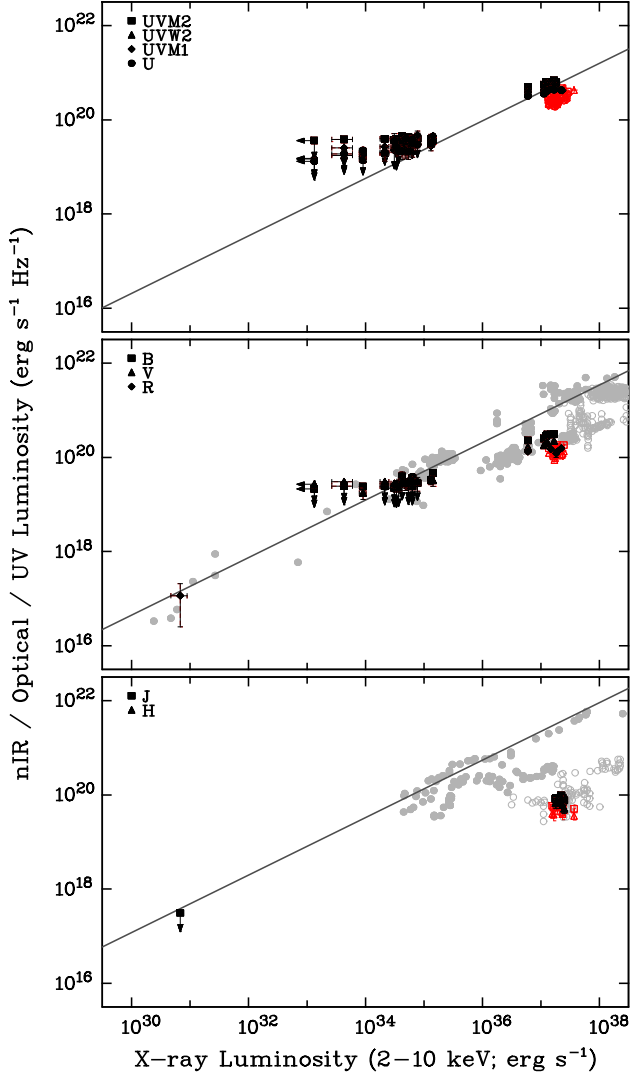}
\caption{Correlation between the nIR/optical/UV and X-ray luminosity during the outburst of \maxinos. 
The \maxi luminosities are calculated for a source distance of 6~kpc. 
The black solid symbols indicate the early times of the outburst when the radio emission in \maxi had not dropped yet (up to day~8 after the outburst onset) and the late times when the source is almost in or in quiescence ($>135$~days), while the red open symbols indicate the times in between. 
The light grey symbols are for other BHXBs during the hard state (solid symbols) and soft state (open symbols) as presented by \citet{russell2006} and \citet{russell2007}. 
The solid lines show the best-fit hard state nIR/optical-X-ray correlation from \citet{russell2006}.}
\label{fig:optxraycor}
\end{center}
\end{figure}

While the radio emission is produced in the jet and the X-ray emission in the accretion disk and/or the corona, the nature of the UV, optical and nIR emission is less unambiguous. 
Most likely the UV emission is also produced in the accretion disk, either intrinsically or by reprocessing of X-rays, but in the optical and nIR there may also be a contribution from the jet. 
In Section~\ref{sec:modresults} we have modeled the full SED from radio to X-ray frequencies with a combination of a broken power law and a physically motivated irradiated disk model. 
From these spectral fits we have concluded that a jet contribution at nIR wavelengths is not necessary. 
This conclusion, however, is model dependent and here we look deeper into this issue by investigating the correlations between the X-ray emission and the nIR, optical and UV emission. 

A correlation between the nIR/optical and X-ray luminosity was first found for GX\,339$-$4 in the hard state by \citet{homan2005a}, with a power-law slope of $\sim0.5$. 
\citet{russell2006} has shown for a large sample of BHXBs in the hard state that there is such a correlation with a power-law slope of $\sim0.6$ over eight orders of magnitude in X-ray luminosity. 
A correlation with such a power-law slope can be expected both in the case of X-ray reprocessing \citep[][]{vanparadijs1994} and jet emission at nIR/optical wavelengths. 
\citet{russell2006} have also shown that when these sources are in a soft state, the nIR and, in some cases, the optical emission is significantly suppressed. 
In Figure~\ref{fig:optxraycor} we show the correlation between the UV/optical/nIR and X-ray luminosities for \maxinos. 
Similar to Figure~\ref{fig:radioxraycor} we have made a distinction between the early outburst before the radio emission strongly decreases (day~8 after the outburst onset; black solid symbols), the second phase of the outburst (red open symbols), and the very late times when the source has reached significantly lower luminosities in the radio, optical and X-rays (black solid symbols). 
We have also included the optical/nIR detections when the source is in quiescence, using the r-band magnitude at $\sim1.5$~years after the outburst onset from \citet{kong2012}, the J-band magnitude at $\sim2.9$~years from this paper, and the quiescent X-ray flux at $\sim1.8$~years reported in \citet{homan2013}. 
From the late-time magnitudes we have subtracted a contribution from the companion star of the black hole. 
It has been proposed \citep{kuulkers2013,kong2012} that this is an M2 or M5 dwarf star, which have an absolute V-band magnitude of 10 and 11.8, respectively \citep{kong2012}. 
We have converted these absolute magnitudes to apparent magnitudes in the r- and J-band by correcting for the typical color of these dwarf stars \citep{johnson1965} and galactic extinction, and by adopting a source distance of 6~kpc. 
The resulting r-band magnitude of the companion star is 24.3 (M2) or 25.8 (M5), and in the J-band it is 21.1 (M2) or 22.0 (M5). 
We have corrected the very-late time fluxes for these ranges in magnitudes, resulting in the large uncertainties displayed in Figure~\ref{fig:optxraycor}.

For comparison we plot in Figure~\ref{fig:optxraycor} the optical/nIR-X-ray relations for the samples presented in \citet{russell2006} and \citet{russell2007}, although here we show the luminosity divided by observing frequency for optical/nIR frequencies, i.e. the measured flux scaled by the source distance. 
For the other BHXBs the solid symbols indicate when they are in a hard state, while the open symbols are for the soft state. 
The top panel of Figure~\ref{fig:optxraycor} shows that the UV emission is consistent with the hard state optical-X-ray correlation \citep[][]{russell2006} throughout the outburst. 
In the middle panel it can be seen that \maxi resides in the same region of the correlation as several other hard state BHXBs in the optical, both early in the outburst before the radio emission is quenched and the late times when the source is going towards or is in quiescence. 
The data corresponding to times after the quenching of the radio emission lay below the correlation of hard state BHXBs, but they are consistent with other sources in the soft state. 
The bottom panel displays different behavior at nIR wavelengths: the data are well below the correlation for other hard state sources throughout the main outburst (no late-time nIR observations are available), but consistent with some BHXBs in the soft state. 
This suggests that at the time of our observations there is no significant contribution of the jet to the nIR emission, which is consistent with our conclusions from the broadband modeling in Section~\ref{sec:modresults}. 
Given the detection of a spectral break at radio frequencies the lack of nIR jet emission is not surprising. 
We note, however, that our nIR observations started $\sim4$~days after the outburst onset, and our earliest detection of a break was at $\sim6$~days, so there is a possibility that the nIR jet emission quenched much earlier than at radio wavelengths. 
This is consistent with what has been found in other sources, namely that the nIR emission from the jet drops as soon as a source enters the HIMS \citep[e.g.][]{homan2005a,coriat2009,russell2010} while the radio flux quenches later \citep[e.g.][]{fender2004}, caused by the jet spectral break shifting to lower frequencies.

\section{Conclusions}\label{sec:conclusions}

In this paper we have presented the results of our observing campaign of \maxi across the electromagnetic spectrum, resulting in one of the richest broadband data sets of a BHXB outburst with observations at radio, sub-millimeter, nIR, optical, UV and X-ray frequencies, from 1~day up to 3~years after the outburst onset. 
We have modeled SEDs from radio to X-ray frequencies at several epochs with a broken power law plus a physical irradiated disk model, and we have presented the radio-X-ray and nIR/optical/UV-X-ray correlations for \maxi in the context of other BHXBs. 

We have found changes in the radio brightness and spectrum which are correlated with X-ray state changes, suggesting a link between the radio jet and the accretion flow that is dominating the X-ray emission. 
We have shown this based on our broadband SED modeling and by investigating the correlation between radio and X-ray luminosity. 
During the first weeks after the outburst onset \maxi moved away from the hard state into intermediate X-ray spectral states, based on spectral and timing behavior at X-ray energies. 
This was further evidenced by (1) the increasing disk temperature, steepening photon index and decreasing disk inner radius we have found in our SED modeling, (2) the quenching of the radio emission coming from the jet, and (3) the deviations from the radio-X-ray and nIR/optical/UV-X-ray correlations of BHXBs in the hard state. 

The broadband radio light curves and spectra support the conclusions from \citet{paragi2013}, based on VLBI observations, that there was no major ejection event. 
Our radio observations also show the presence of a variable spectral break at radio frequencies during the HIMS and SIMS, significantly lower in frequency than typically found in hard state BHXBs. 
We have investigated the nIR/optical/UV-X-ray correlations and concluded that there is no significant contribution from the jet at nIR wavelengths, consistent with our broadband modeling. 
Taken with the results of \citet[][]{russell2013b} and \citet[][]{corbel2013b}, this suggests that the spectral break, which lies at infrared frequencies during the hard state, evolves down in frequency during the HIMS and SIMS until it passes through the radio band.

Our findings for \maxi demonstrate that broadband observations covering several radio and nIR/optical/UV frequencies at a high cadence during BHXB outbursts are important for a better understanding of the jet and disk in these sources. On top of these observations, denser sampling at sub-millimeter and mid-infrared frequencies are crucial to accurately determine the evolution of the spectral break and further our understanding of the jet physics in these systems.

\section*{Acknowledgments}
We greatly appreciate the support from the VLA, WSRT, APEX, ATCA, GMRT, SMARTS, IAC80, 1.23m CAHA, BOOTES-2 and BOOTES-3 telescopes in their help with scheduling and obtaining these observations. 
The National Radio Astronomy Observatory is operated by Associated Universities, Inc., 
under cooperative agreement with the National Science Foundation. 
The WSRT is operated by ASTRON (Netherlands Institute for Radio Astronomy) 
with support from the Netherlands foundation for Scientific Research. 
APEX is a collaboration between the Max-Plank-Institut f\"{u}r Radioastronomie, the European Southern Observatory and the Onsala Space Observatory. 
The ATCA is funded by the Commonwealth of Australia for operation as a National Facility managed by CSIRO. 
The GMRT is operated by the National Center for Radio Astrophysics of the Tata Institute of Fundamental Research. 
The CTIO 1.3m~telescope is operated by the SMARTS consortium. 
The Centro Astron\'omico Hispano Alem\'an (CAHA) at Calar Alto is operated jointly by the Max-Planck Institut  f\"ur Astronomie and the Instituto de Astrof\'{\i}sica de Andaluc\'{\i}a (CSIC). 
The Liverpool Telescope is operated on the island of La Palma by Liverpool John Moores University in the Spanish Observatorio del Roque de los Muchachos of the Instituto de Astrofisica de Canarias with financial support from the UK Science and Technology Facilities Council. 
The IAC80 is operated on the island of Tenerife by the IAC in the Spanish Observatorio del Teide. 
The MAXI/GSC data are provided by RIKEN, JAXA and the MAXI team. 
This research has made use of data obtained from the High Energy Astrophysics Science Archive Research Center (HEASARC), provided by NASA's Goddard Space Flight Center.

AJvdH and RAMJW acknowledge support from the European Research Council via Advanced Investigator Grant no. 247295. 
PAC and JCAMJ acknowledge support from the Australian Research Council's {\it Discovery Projects} funding scheme under grant DP120102393. 
DMR acknowledges support from a Marie Curie Intra European Fellowship within the 7th European Community Framework Programme under contract no. IEF 274805. 
AdUP acknowledges  support by the European Commission under the Marie Curie Career Integration Grant programme (FP7-PEOPLE-2012-CIG  322307), and the Dark Cosmology Centre, funded by the DNRF. 
This work was supported by he Unidad Asociada IAA-CSIC at the group of planetary science of  ETSI-UPV/EHU, by the Ikerbasque Foundation for Science, and by Spanish research programs AYA2012-39362-C02-02, AYA2011-24780/ESP, AYA2009-14000-C03-01/ESP and AYA2010-21887-C04-01. 
TB acknowledges support from grant PRIN-INAF 2012. 
RW acknowledges support from the European Research Council via a Starting Grant.

\bibliographystyle{mn2e}
\bibliography{references}

\begin{thebibliography}{95}
\expandafter\ifx\csname natexlab\endcsname\relax\def\natexlab#1{#1}\fi

\bibitem[{{Belloni}(2010)}]{belloni2010}
{Belloni} T., ed., 2010, Lecture Notes in Physics, Berlin Springer Verlag, Vol.
  794, {The Jet Paradigm}

\bibitem[{{Belloni}, {Motta} \& {Mu{\~n}oz-Darias}(2010){Belloni}, {Motta}, \&
  {Mu{\~n}oz-Darias}}]{belloniatel2927}
{Belloni} T.~M., {Motta} S., {Mu{\~n}oz-Darias} T., 2010, The Astronomer's
  Telegram, 2927, 1

\bibitem[{{Belloni}, {Mu{\~n}oz-Darias} \& {Kuulkers}(2010){Belloni},
  {Mu{\~n}oz-Darias}, \& {Kuulkers}}]{belloniatel2926}
{Belloni} T.~M., {Mu{\~n}oz-Darias} T., {Kuulkers} E., 2010, The Astronomer's
  Telegram, 2926, 1

\bibitem[{{Breeveld} {et~al}\mbox{.}(2010){Breeveld}, {Curran}, {Hoversten},
  {Koch}, {Landsman}, {Marshall}, {Page}, {Poole}, {Roming}, {Smith}, {Still},
  {Yershov}, {Blustin}, {Brown}, {Gronwall}, {Holland}, {Kuin}, {McGowan},
  {Rosen}, {Boyd}, {Broos}, {Carter}, {Chester}, {Hancock}, {Huckle}, {Immler},
  {Ivanushkina}, {Kennedy}, {Mason}, {Morgan}, {Oates}, {de Pasquale},
  {Schady}, {Siegel}, \& {vanden Berk}}]{breeveld2010}
{Breeveld} A.~A. {et~al.}, 2010, \mnras, 406, 1687

\bibitem[{{Casella}, {Belloni} \& {Stella}(2005){Casella}, {Belloni}, \&
  {Stella}}]{casella2005}
{Casella} P., {Belloni} T., {Stella} L., 2005, \apj, 629, 403

\bibitem[{{Castro-Tirado} {et~al}\mbox{.}(1999){Castro-Tirado}, {Sold{\'a}n},
  {Bernas}, {P{\'a}ta}, {Rezek}, {Hudec}, {Mateo Sanguino}, {de La Morena},
  {Bern{\'a}}, {Rodr{\'{\i}}guez}, {Pe{\~n}a}, {Gorosabel}, {M{\'a}s-Hesse}, \&
  {Gim{\'e}nez}}]{castrotirado1999}
{Castro-Tirado} A.~J. {et~al.}, 1999, \aaps, 138, 583

\bibitem[{{Corbel} {et~al}\mbox{.}(2013{\natexlab{a}}){Corbel}, {Aussel},
  {Broderick}, {Chanial}, {Coriat}, {Maury}, {Buxton}, {Tomsick}, {Tzioumis},
  {Markoff}, {Rodriguez}, {Bailyn}, {Brocksopp}, {Fender}, {Petrucci},
  {Cadolle-Bel}, {Calvelo}, \& {Harvey-Smith}}]{corbel2013b}
{Corbel} S. {et~al.}, 2013{\natexlab{a}}, \mnras, 431, L107

\bibitem[{{Corbel} {et~al}\mbox{.}(2013{\natexlab{b}}){Corbel}, {Coriat},
  {Brocksopp}, {Tzioumis}, {Fender}, {Tomsick}, {Buxton}, \&
  {Bailyn}}]{corbel2013a}
{Corbel} S., {Coriat} M., {Brocksopp} C., {Tzioumis} A.~K., {Fender} R.~P.,
  {Tomsick} J.~A., {Buxton} M.~M., {Bailyn} C.~D., 2013{\natexlab{b}}, \mnras,
  428, 2500

\bibitem[{{Corbel} \& {Fender}(2002)}]{corbel2002}
{Corbel} S., {Fender} R.~P., 2002, \apjl, 573, L35

\bibitem[{{Corbel}, {Koerding} \& {Kaaret}(2008){Corbel}, {Koerding}, \&
  {Kaaret}}]{corbel2008}
{Corbel} S., {Koerding} E., {Kaaret} P., 2008, \mnras, 389, 1697

\bibitem[{{Corbel} {et~al}\mbox{.}(2003){Corbel}, {Nowak}, {Fender},
  {Tzioumis}, \& {Markoff}}]{corbel2003}
{Corbel} S., {Nowak} M.~A., {Fender} R.~P., {Tzioumis} A.~K., {Markoff} S.,
  2003, \aap, 400, 1007

\bibitem[{{Coriat} {et~al}\mbox{.}(2009){Coriat}, {Corbel}, {Buxton}, {Bailyn},
  {Tomsick}, {K{\"o}rding}, \& {Kalemci}}]{coriat2009}
{Coriat} M., {Corbel} S., {Buxton} M.~M., {Bailyn} C.~D., {Tomsick} J.~A.,
  {K{\"o}rding} E., {Kalemci} E., 2009, \mnras, 400, 123

\bibitem[{{Coriat} {et~al}\mbox{.}(2011){Coriat}, {Corbel}, {Prat},
  {Miller-Jones}, {Cseh}, {Tzioumis}, {Brocksopp}, {Rodriguez}, {Fender}, \&
  {Sivakoff}}]{coriat2011}
{Coriat} M. {et~al.}, 2011, \mnras, 414, 677

\bibitem[{{de Ugarte Postigo} {et~al}\mbox{.}(2010{\natexlab{a}}){de Ugarte
  Postigo}, {Flores}, {Wiersema}, {Thoene}, {Fynbo}, \&
  {Goldoni}}]{deugartepostigogcn11307}
{de Ugarte Postigo} A., {Flores} H., {Wiersema} K., {Thoene} C.~C., {Fynbo}
  J.~P.~U., {Goldoni} P., 2010{\natexlab{a}}, GRB Coordinates Network, Circular
  Service, 11307, 1 (2010), 1307, 1

\bibitem[{{de Ugarte Postigo} {et~al}\mbox{.}(2010{\natexlab{b}}){de Ugarte
  Postigo}, {Lundrgren}, {Wyrowski}, {Thoene}, {Castro-Tirado}, {Gorosabel}, \&
  {Jelinek}}]{deugartepostigogcn11304}
{de Ugarte Postigo} A., {Lundrgren} A., {Wyrowski} F., {Thoene} C.~C.,
  {Castro-Tirado} A.~J., {Gorosabel} J., {Jelinek} M., 2010{\natexlab{b}}, GRB
  Coordinates Network, Circular Service, 11304, 1 (2010), 1304, 1

\bibitem[{{Evans} {et~al}\mbox{.}(2009){Evans}, {Beardmore}, {Page}, {Osborne},
  {O'Brien}, {Willingale}, {Starling}, {Burrows}, {Godet}, {Vetere}, {Racusin},
  {Goad}, {Wiersema}, {Angelini}, {Capalbi}, {Chincarini}, {Gehrels}, {Kennea},
  {Margutti}, {Morris}, {Mountford}, {Pagani}, {Perri}, {Romano}, \&
  {Tanvir}}]{evans2009}
{Evans} P.~A. {et~al.}, 2009, \mnras, 397, 1177

\bibitem[{{Fender}(2010)}]{fender2010}
{Fender} R., 2010, in Lecture Notes in Physics, Berlin Springer Verlag, Vol.
  794, Lecture Notes in Physics, Berlin Springer Verlag, {Belloni} T., ed., p.
  115

\bibitem[{{Fender}(2001)}]{fender2001}
{Fender} R.~P., 2001, \mnras, 322, 31

\bibitem[{{Fender}, {Belloni} \& {Gallo}(2004){Fender}, {Belloni}, \&
  {Gallo}}]{fender2004}
{Fender} R.~P., {Belloni} T.~M., {Gallo} E., 2004, \mnras, 355, 1105

\bibitem[{{Fender}, {Homan} \& {Belloni}(2009){Fender}, {Homan}, \&
  {Belloni}}]{fender2009}
{Fender} R.~P., {Homan} J., {Belloni} T.~M., 2009, \mnras, 396, 1370

\bibitem[{{Fukugita} {et~al}\mbox{.}(1996){Fukugita}, {Ichikawa}, {Gunn},
  {Doi}, {Shimasaku}, \& {Schneider}}]{fukugita1996}
{Fukugita} M., {Ichikawa} T., {Gunn} J.~E., {Doi} M., {Shimasaku} K.,
  {Schneider} D.~P., 1996, \aj, 111, 1748

\bibitem[{{Fukugita}, {Shimasaku} \& {Ichikawa}(1995){Fukugita}, {Shimasaku},
  \& {Ichikawa}}]{fukugita1995}
{Fukugita} M., {Shimasaku} K., {Ichikawa} T., 1995, \pasp, 107, 945

\bibitem[{{Gallo}(2010)}]{gallo2010}
{Gallo} E., 2010, in Lecture Notes in Physics, Berlin Springer Verlag, Vol.
  794, Lecture Notes in Physics, Berlin Springer Verlag, {Belloni} T., ed.,
  p.~85

\bibitem[{{Gallo}, {Fender} \& {Pooley}(2003){Gallo}, {Fender}, \&
  {Pooley}}]{gallo2003}
{Gallo} E., {Fender} R.~P., {Pooley} G.~G., 2003, \mnras, 344, 60

\bibitem[{{Gallo}, {Miller} \& {Fender}(2012){Gallo}, {Miller}, \&
  {Fender}}]{gallo2012}
{Gallo} E., {Miller} B.~P., {Fender} R., 2012, \mnras, 423, 590

\bibitem[{{Gandhi} {et~al}\mbox{.}(2011){Gandhi}, {Blain}, {Russell},
  {Casella}, {Malzac}, {Corbel}, {D'Avanzo}, {Lewis}, {Markoff}, {Cadolle Bel},
  {Goldoni}, {Wachter}, {Khangulyan}, \& {Mainzer}}]{gandhi2011}
{Gandhi} P. {et~al.}, 2011, \apjl, 740, L13

\bibitem[{{Gierli{\'n}ski}, {Done} \& {Page}(2008){Gierli{\'n}ski}, {Done}, \&
  {Page}}]{gierlinski2008}
{Gierli{\'n}ski} M., {Done} C., {Page} K., 2008, \mnras, 388, 753

\bibitem[{{Gierli{\'n}ski}, {Done} \& {Page}(2009){Gierli{\'n}ski}, {Done}, \&
  {Page}}]{gierlinski2009}
{Gierli{\'n}ski} M., {Done} C., {Page} K., 2009, \mnras, 392, 1106

\bibitem[{{Gilfanov}(2010)}]{gilfanov2010}
{Gilfanov} M., 2010, in Lecture Notes in Physics, Berlin Springer Verlag, Vol.
  794, Lecture Notes in Physics, Berlin Springer Verlag, {Belloni} T., ed.,
  p.~17

\bibitem[{{Hannikainen} {et~al}\mbox{.}(1998){Hannikainen}, {Hunstead},
  {Campbell-Wilson}, \& {Sood}}]{hannikainen1998}
{Hannikainen} D.~C., {Hunstead} R.~W., {Campbell-Wilson} D., {Sood} R.~K.,
  1998, \aap, 337, 460

\bibitem[{{Homan} \& {Belloni}(2005)}]{homan2005b}
{Homan} J., {Belloni} T., 2005, \apss, 300, 107

\bibitem[{{Homan} {et~al}\mbox{.}(2005){Homan}, {Buxton}, {Markoff}, {Bailyn},
  {Nespoli}, \& {Belloni}}]{homan2005a}
{Homan} J., {Buxton} M., {Markoff} S., {Bailyn} C.~D., {Nespoli} E., {Belloni}
  T., 2005, \apj, 624, 295

\bibitem[{{Homan} {et~al}\mbox{.}(2013){Homan}, {Fridriksson}, {Jonker},
  {Russell}, {Gallo}, {Kuulkers}, {Rea}, \& {Altamirano}}]{homan2013}
{Homan} J., {Fridriksson} J.~K., {Jonker} P.~G., {Russell} D.~M., {Gallo} E.,
  {Kuulkers} E., {Rea} N., {Altamirano} D., 2013, ArXiv e-prints

\bibitem[{{Homan} {et~al}\mbox{.}(2001){Homan}, {Wijnands}, {van der Klis},
  {Belloni}, {van Paradijs}, {Klein-Wolt}, {Fender}, \&
  {M{\'e}ndez}}]{homan2001}
{Homan} J., {Wijnands} R., {van der Klis} M., {Belloni} T., {van Paradijs} J.,
  {Klein-Wolt} M., {Fender} R., {M{\'e}ndez} M., 2001, \apjs, 132, 377

\bibitem[{{Jelinek} {et~al}\mbox{.}(2010){Jelinek}, {Zurita}, {Visus},
  {Papics}, {Kubanek}, {Sabau-Graziati}, {de Ugarte Postigo}, {Cunniffe},
  {Gorosabel}, \& {Castro-Tirado}}]{jelinekgcn11301}
{Jelinek} M. {et~al.}, 2010, GRB Coordinates Network, Circular Service, 11301,
  1 (2010), 1301, 1

\bibitem[{{Johnson}(1965)}]{johnson1965}
{Johnson} H.~L., 1965, \apj, 141, 170

\bibitem[{{Jonker} {et~al}\mbox{.}(2012){Jonker}, {Miller-Jones}, {Homan},
  {Tomsick}, {Fender}, {Kaaret}, {Markoff}, \& {Gallo}}]{jonker2012}
{Jonker} P.~G., {Miller-Jones} J.~C.~A., {Homan} J., {Tomsick} J., {Fender}
  R.~P., {Kaaret} P., {Markoff} S., {Gallo} E., 2012, \mnras, 423, 3308

\bibitem[{{Jordi}, {Grebel} \& {Ammon}(2006){Jordi}, {Grebel}, \&
  {Ammon}}]{jordi2006}
{Jordi} K., {Grebel} E.~K., {Ammon} K., 2006, \aap, 460, 339

\bibitem[{{Kalamkar} {et~al}\mbox{.}(2011){Kalamkar}, {Homan}, {Altamirano},
  {van der Klis}, {Casella}, \& {Linares}}]{kalamkar2011}
{Kalamkar} M., {Homan} J., {Altamirano} D., {van der Klis} M., {Casella} P.,
  {Linares} M., 2011, \apjl, 731, L2

\bibitem[{{Kalamkar} {et~al}\mbox{.}(2010){Kalamkar}, {Yang}, {Altamirano},
  {Casella}, {Linares}, {Patruno}, {Armas-Padilla}, {Cavecchi}, {Degenaar},
  {Russell}, {Kaur}, {van der Klis}, {Watts}, {Wijnands}, \&
  {Rea}}]{kalamkaratel2881}
{Kalamkar} M. {et~al.}, 2010, The Astronomer's Telegram, 2881, 1

\bibitem[{{Kalberla} {et~al}\mbox{.}(2005){Kalberla}, {Burton}, {Hartmann},
  {Arnal}, {Bajaja}, {Morras}, \& {P{\"o}ppel}}]{kalberla2005}
{Kalberla} P.~M.~W., {Burton} W.~B., {Hartmann} D., {Arnal} E.~M., {Bajaja} E.,
  {Morras} R., {P{\"o}ppel} W.~G.~L., 2005, \aap, 440, 775

\bibitem[{{Kann}(2010)}]{kanngcn11299}
{Kann} D.~A., 2010, GRB Coordinates Network, Circular Service, 11299, 1 (2010),
  1299, 1

\bibitem[{{Kaur} {et~al}\mbox{.}(2012){Kaur}, {Kaper}, {Ellerbroek}, {Russell},
  {Altamirano}, {Wijnands}, {Yang}, {D'Avanzo}, {de Ugarte Postigo}, {Flores},
  {Fynbo}, {Goldoni}, {Th{\"o}ne}, {van der Horst}, {van der Klis},
  {Kouveliotou}, {Wiersema}, \& {Kuulkers}}]{kaur2012}
{Kaur} R. {et~al.}, 2012, \apjl, 746, L23

\bibitem[{{Kennea} {et~al}\mbox{.}(2010){Kennea}, {Krimm}, {Mangano}, {Curran},
  {Romano}, {Evans}, \& {Burrows}}]{kenneaatel2877}
{Kennea} J.~A., {Krimm} H., {Mangano} V., {Curran} P., {Romano} P., {Evans} P.,
  {Burrows} D.~N., 2010, The Astronomer's Telegram, 2877, 1

\bibitem[{{Kennea} {et~al}\mbox{.}(2011){Kennea}, {Romano}, {Mangano},
  {Beardmore}, {Evans}, {Curran}, {Krimm}, {Markwardt}, \&
  {Yamaoka}}]{kennea2011}
{Kennea} J.~A. {et~al.}, 2011, \apj, 736, 22

\bibitem[{{Kong}(2012)}]{kong2012}
{Kong} A.~K.~H., 2012, \apjl, 760, L27

\bibitem[{{Kuulkers} {et~al}\mbox{.}(1999){Kuulkers}, {Fender}, {Spencer},
  {Davis}, \& {Morison}}]{kuulkers1999}
{Kuulkers} E., {Fender} R.~P., {Spencer} R.~E., {Davis} R.~J., {Morison} I.,
  1999, \mnras, 306, 919

\bibitem[{{Kuulkers} {et~al}\mbox{.}(2010){Kuulkers}, {Ibarra}, {Pollock},
  {Parmar}, {Chevenez}, {Kouveliotou}, {van der Horst}, {Paragi}, {Granot}, \&
  {Taylor}}]{kuulkersatel2912}
{Kuulkers} E. {et~al.}, 2010, The Astronomer's Telegram, 2912, 1

\bibitem[{{Kuulkers} {et~al}\mbox{.}(2013){Kuulkers}, {Kouveliotou}, {Belloni},
  {Cadolle Bel}, {Chenevez}, {D{\'{\i}}az Trigo}, {Homan}, {Ibarra}, {Kennea},
  {Mu{\~n}oz-Darias}, {Ness}, {Parmar}, {Pollock}, {van den Heuvel}, \& {van
  der Horst}}]{kuulkers2013}
{Kuulkers} E. {et~al.}, 2013, \aap, 552, A32

\bibitem[{{Kuulkers} {et~al}\mbox{.}(2012){Kuulkers}, {Kouveliotou}, {van der
  Horst}, {Belloni}, {Chenevez}, {Ibarra}, {Munoz-Darias}, {Bazzano}, {Cadolle
  Bel}, {De Cesare}, {Diaz Trigo}, {Jourdain}, {Lubinski}, {Natalucci}, {Ness},
  {Parmar}, {Pollock}, {Rodriguez}, {Roques}, {Sanchez-Fernandez}, {Ubertini},
  \& {Winkler}}]{kuulkers2012}
{Kuulkers} E. {et~al.}, 2012, in The First Year of MAXI: Monitoring variable
  X-ray sources -- 4th International MAXI Workshop, {Mihara} T. \&~{Serino} M.,
  ed., Vol. IPCR CR-127, 81

\bibitem[{{Lundgren} {et~al}\mbox{.}(2010){Lundgren}, {Rabanus}, {G{\"u}sten},
  {Menten}, {de Zeeuw}, {Olofsson}, {Kaufer}, {Nyman}, {Bergman}, {De Breuck},
  {Wyrowski}, {Agurto}, {Azagra}, {Dumke}, {Mac-Auliffe}, {Martinez},
  {Montenegro}, {Muders}, {Reveret}, {Risacher}, {Parra}, {Siringo}, \&
  {Wieching}}]{lundgren2010}
{Lundgren} A. {et~al.}, 2010, in Society of Photo-Optical Instrumentation
  Engineers (SPIE) Conference Series, Vol. 7737, Society of Photo-Optical
  Instrumentation Engineers (SPIE) Conference Series

\bibitem[{{Malzac}, {Dumont} \& {Mouchet}(2005){Malzac}, {Dumont}, \&
  {Mouchet}}]{malzac2005}
{Malzac} J., {Dumont} A.~M., {Mouchet} M., 2005, \aap, 430, 761

\bibitem[{{Mangano} {et~al}\mbox{.}(2010){Mangano}, {Hoversten}, {Markwardt},
  {Sbarufatti}, {Starling}, \& {Ukwatta}}]{manganogcn11296}
{Mangano} V., {Hoversten} E.~A., {Markwardt} C.~B., {Sbarufatti} B., {Starling}
  R.~L.~C., {Ukwatta} T.~N., 2010, GRB Coordinates Network, Circular Service,
  11296, 1 (2010), 1296, 1

\bibitem[{{Markoff} \& {Nowak}(2004)}]{markoff2004}
{Markoff} S., {Nowak} M.~A., 2004, \apj, 609, 972

\bibitem[{{Marshall}(2010)}]{marshallgcn11298}
{Marshall} F.~E., 2010, GRB Coordinates Network, Circular Service, 11298, 1
  (2010), 1298, 1

\bibitem[{{McClintock} \& {Remillard}(2006)}]{mcclintock2006}
{McClintock} J.~E., {Remillard} R.~A., 2006, {Black hole binaries}, {Lewin}
  W.~H.~G., {van der Klis} M., eds., pp. 157--213

\bibitem[{{Miller}, {Homan} \& {Miniutti}(2006){Miller}, {Homan}, \&
  {Miniutti}}]{miller2006}
{Miller} J.~M., {Homan} J., {Miniutti} G., 2006, \apjl, 652, L113

\bibitem[{{Miller-Jones} {et~al}\mbox{.}(2011){Miller-Jones}, {Madej},
  {Jonker}, {Homan}, {Ratti}, \& {Torres}}]{millerjonesatel3358}
{Miller-Jones} J.~C.~A., {Madej} O.~K., {Jonker} P.~G., {Homan} J., {Ratti}
  E.~M., {Torres} M.~A.~P., 2011, The Astronomer's Telegram, 3358, 1

\bibitem[{{Miller-Jones} {et~al}\mbox{.}(2012){Miller-Jones}, {Sivakoff},
  {Altamirano}, {Coriat}, {Corbel}, {Dhawan}, {Krimm}, {Remillard}, {Rupen},
  {Russell}, {Fender}, {Heinz}, {K{\"o}rding}, {Maitra}, {Markoff}, {Migliari},
  {Sarazin}, \& {Tudose}}]{millerjones2012}
{Miller-Jones} J.~C.~A. {et~al.}, 2012, \mnras, 421, 468

\bibitem[{{Mirabel} {et~al}\mbox{.}(1992){Mirabel}, {Rodriguez}, {Cordier},
  {Paul}, \& {Lebrun}}]{mirabel1992}
{Mirabel} I.~F., {Rodriguez} L.~F., {Cordier} B., {Paul} J., {Lebrun} F., 1992,
  \nat, 358, 215

\bibitem[{{Monet} {et~al}\mbox{.}(2003){Monet}, {Levine}, {Canzian}, {Ables},
  {Bird}, {Dahn}, {Guetter}, {Harris}, {Henden}, {Leggett}, {Levison},
  {Luginbuhl}, {Martini}, {Monet}, {Munn}, {Pier}, {Rhodes}, {Riepe}, {Sell},
  {Stone}, {Vrba}, {Walker}, {Westerhout}, {Brucato}, {Reid}, {Schoening},
  {Hartley}, {Read}, \& {Tritton}}]{monet2003}
{Monet} D.~G. {et~al.}, 2003, \aj, 125, 984

\bibitem[{{Motta} {et~al}\mbox{.}(2011){Motta}, {Mu{\~n}oz-Darias}, {Casella},
  {Belloni}, \& {Homan}}]{motta2011}
{Motta} S., {Mu{\~n}oz-Darias} T., {Casella} P., {Belloni} T., {Homan} J.,
  2011, \mnras, 418, 2292

\bibitem[{{Mu{\~n}oz-Darias} {et~al}\mbox{.}(2011){Mu{\~n}oz-Darias}, {Motta},
  {Stiele}, \& {Belloni}}]{munozdarias2011}
{Mu{\~n}oz-Darias} T., {Motta} S., {Stiele} H., {Belloni} T.~M., 2011, \mnras,
  415, 292

\bibitem[{{Mu{\~n}oz-Darias} {et~al}\mbox{.}(2010){Mu{\~n}oz-Darias}, {Stiele},
  {Belloni}, \& {Motta}}]{munozdariasatel2999}
{Mu{\~n}oz-Darias} T., {Stiele} H., {Belloni} T.~M., {Motta} S., 2010, The
  Astronomer's Telegram, 2999, 1

\bibitem[{{Negoro} {et~al}\mbox{.}(2010){Negoro}, {Yamaoka}, {Nakahira},
  {Kawasaki}, {Ueno}, {Tomida}, {Kohama}, {Ishikawa}, {Mihara}, {Nakagawa},
  {Sugizaki}, {Serino}, {Yamamoto}, {Sootome}, {Matsuoka}, {Kawai}, {Morii},
  {Sugimori}, {Usui}, {Yoshida}, {Tsunemi}, {Kimura}, {Nakajima}, {Ozawa},
  {Suwa}, {Ueda}, {Isobe}, {Eguchi}, {Hiroi}, {Daikyuji}, {Uzawa}, {Yamazaki},
  \& {Matsumura}}]{negoroatel2873}
{Negoro} H. {et~al.}, 2010, The Astronomer's Telegram, 2873, 1

\bibitem[{{Paragi} {et~al}\mbox{.}(2013){Paragi}, {van der Horst}, {Belloni},
  {Miller-Jones}, {Linford}, {Taylor}, {Yang}, {Garrett}, {Granot},
  {Kouveliotou}, {Kuulkers}, \& {Wijers}}]{paragi2013}
{Paragi} Z. {et~al.}, 2013, \mnras, 432, 1319

\bibitem[{{Paragi} {et~al}\mbox{.}(2010){Paragi}, {van der Horst}, {Granot},
  {Taylor}, {Kouveliotou}, {Garrett}, {Wijers}, {Ramirez-Ruiz}, {Kuulkers},
  {Gehrels}, \& {Woods}}]{paragiatel2906}
{Paragi} Z. {et~al.}, 2010, The Astronomer's Telegram, 2906, 1

\bibitem[{{Poole} {et~al}\mbox{.}(2008){Poole}, {Breeveld}, {Page}, {Landsman},
  {Holland}, {Roming}, {Kuin}, {Brown}, {Gronwall}, {Hunsberger}, {Koch},
  {Mason}, {Schady}, {vanden Berk}, {Blustin}, {Boyd}, {Broos}, {Carter},
  {Chester}, {Cucchiara}, {Hancock}, {Huckle}, {Immler}, {Ivanushkina},
  {Kennedy}, {Marshall}, {Morgan}, {Pandey}, {de Pasquale}, {Smith}, \&
  {Still}}]{poole2008}
{Poole} T.~S. {et~al.}, 2008, \mnras, 383, 627

\bibitem[{{Ratti} {et~al}\mbox{.}(2012){Ratti}, {Jonker}, {Miller-Jones},
  {Torres}, {Homan}, {Markoff}, {Tomsick}, {Kaaret}, {Wijnands}, {Gallo},
  {{\"O}zel}, {Steeghs}, \& {Fender}}]{ratti2012}
{Ratti} E.~M. {et~al.}, 2012, \mnras, 423, 2656

\bibitem[{{Rushton} {et~al}\mbox{.}(2012){Rushton}, {Miller-Jones}, {Campana},
  {Evangelista}, {Paragi}, {Maccarone}, {Pooley}, {Tudose}, {Fender},
  {Spencer}, \& {Dhawan}}]{rushton2012}
{Rushton} A. {et~al.}, 2012, \mnras, 419, 3194

\bibitem[{{Russell} {et~al}\mbox{.}(2006){Russell}, {Fender}, {Hynes},
  {Brocksopp}, {Homan}, {Jonker}, \& {Buxton}}]{russell2006}
{Russell} D.~M., {Fender} R.~P., {Hynes} R.~I., {Brocksopp} C., {Homan} J.,
  {Jonker} P.~G., {Buxton} M.~M., 2006, \mnras, 371, 1334

\bibitem[{{Russell} {et~al}\mbox{.}(2010{\natexlab{a}}){Russell}, {Lewis},
  {Bersier}, {Cano}, {Gandhi}, {Patruno}, {Kalamkar}, {Yang}, {Altamirano},
  {Casella}, {Linares}, {Padilla}, {Cavecchi}, {Degenaar}, {Kaur}, {van der
  Klis}, {Watts}, {Wijnands}, \& {Rea}}]{russellatel2884}
{Russell} D.~M. {et~al.}, 2010{\natexlab{a}}, The Astronomer's Telegram, 2884,
  1

\bibitem[{{Russell} {et~al}\mbox{.}(2007){Russell}, {Maccarone}, {K{\"o}rding},
  \& {Homan}}]{russell2007}
{Russell} D.~M., {Maccarone} T.~J., {K{\"o}rding} E.~G., {Homan} J., 2007,
  \mnras, 379, 1401

\bibitem[{{Russell} {et~al}\mbox{.}(2010{\natexlab{b}}){Russell}, {Maitra},
  {Dunn}, \& {Markoff}}]{russell2010}
{Russell} D.~M., {Maitra} D., {Dunn} R.~J.~H., {Markoff} S.,
  2010{\natexlab{b}}, \mnras, 405, 1759

\bibitem[{{Russell} {et~al}\mbox{.}(2013{\natexlab{a}}){Russell}, {Markoff},
  {Casella}, {Cantrell}, {Chatterjee}, {Fender}, {Gallo}, {Gandhi}, {Homan},
  {Maitra}, {Miller-Jones}, {O'Brien}, \& {Shahbaz}}]{russell2013}
{Russell} D.~M. {et~al.}, 2013{\natexlab{a}}, \mnras, 429, 815

\bibitem[{{Russell} {et~al}\mbox{.}(2013{\natexlab{b}}){Russell}, {Russell},
  {Miller-Jones}, {O'Brien}, {Soria}, {Sivakoff}, {Slaven-Blair}, {Lewis},
  {Markoff}, {Homan}, {Altamirano}, {Curran}, {Rupen}, {Belloni}, {Cadolle
  Bel}, {Casella}, {Corbel}, {Dhawan}, {Fender}, {Gallo}, {Gandhi}, {Heinz},
  {K{\"o}rding}, {Krimm}, {Maitra}, {Migliari}, {Remillard}, {Sarazin},
  {Shahbaz}, \& {Tudose}}]{russell2013b}
{Russell} D.~M. {et~al.}, 2013{\natexlab{b}}, \apjl, 768, L35

\bibitem[{{Rykoff} {et~al}\mbox{.}(2007){Rykoff}, {Miller}, {Steeghs}, \&
  {Torres}}]{rykoff2007}
{Rykoff} E.~S., {Miller} J.~M., {Steeghs} D., {Torres} M.~A.~P., 2007, \apj,
  666, 1129

\bibitem[{{Sault}, {Teuben} \& {Wright}(1995){Sault}, {Teuben}, \&
  {Wright}}]{sault1995}
{Sault} R.~J., {Teuben} P.~J., {Wright} M.~C.~H., 1995, in Astronomical Society
  of the Pacific Conference Series, Vol.~77, Astronomical Data Analysis
  Software and Systems IV, {Shaw} R.~A., {Payne} H.~E., {Hayes} J.~J.~E., eds.,
  p. 433

\bibitem[{{Schlegel}, {Finkbeiner} \& {Davis}(1998){Schlegel}, {Finkbeiner}, \&
  {Davis}}]{schlegel1998}
{Schlegel} D.~J., {Finkbeiner} D.~P., {Davis} M., 1998, \apj, 500, 525

\bibitem[{{Shakura} \& {Sunyaev}(1973)}]{shakura1973}
{Shakura} N.~I., {Sunyaev} R.~A., 1973, \aap, 24, 337

\bibitem[{{Shaposhnikov} \& {Yamaoka}(2010)}]{shaposhnikovatel2951}
{Shaposhnikov} N., {Yamaoka} K., 2010, The Astronomer's Telegram, 2951, 1

\bibitem[{{Siringo} {et~al}\mbox{.}(2009){Siringo}, {Kreysa}, {Kov{\'a}cs},
  {Schuller}, {Wei{\ss}}, {Esch}, {Gem{\"u}nd}, {Jethava}, {Lundershausen},
  {Colin}, {G{\"u}sten}, {Menten}, {Beelen}, {Bertoldi}, {Beeman}, \&
  {Haller}}]{siringo2009}
{Siringo} G. {et~al.}, 2009, \aap, 497, 945

\bibitem[{{Skrutskie} {et~al}\mbox{.}(1995){Skrutskie}, {Beichman}, {Capps},
  {Carpenter}, {Chester}, {Cutri}, {Elias}, {Elston}, {Huchra}, {Liebert},
  {Lonsdale}, {Monet}, {Price}, {Schneider}, {Seitzer}, {Stiening}, {Strom}, \&
  {Weinberg}}]{skrutskie1995}
{Skrutskie} M.~F. {et~al.}, 1995, in Bulletin of the American Astronomical
  Society, Vol.~27, American Astronomical Society Meeting Abstracts, p. 1392

\bibitem[{{Soria}(2007)}]{soria2007}
{Soria} R., 2007, \apss, 311, 213

\bibitem[{{Subasavage} {et~al}\mbox{.}(2010){Subasavage}, {Bailyn}, {Smith},
  {Henry}, {Walter}, \& {Buxton}}]{subasavage2010}
{Subasavage} J.~P., {Bailyn} C.~D., {Smith} R.~C., {Henry} T.~J., {Walter}
  F.~M., {Buxton} M.~M., 2010, in Society of Photo-Optical Instrumentation
  Engineers (SPIE) Conference Series, Vol. 7737, Society of Photo-Optical
  Instrumentation Engineers (SPIE) Conference Series

\bibitem[{{Tan}(1991)}]{tan1991}
{Tan} G.~H., 1991, in Astronomical Society of the Pacific Conference Series,
  Vol.~19, IAU Colloq. 131: Radio Interferometry. Theory, Techniques, and
  Applications, {Cornwell} T.~J., {Perley} R.~A., eds., pp. 42--46

\bibitem[{{van der Horst} {et~al}\mbox{.}(2010{\natexlab{a}}){van der Horst},
  {Granot}, {Paragi}, {Kouveliotou}, {Wijers}, \&
  {Ramirez-Ruiz}}]{vanderhorstgcn11309}
{van der Horst} A.~J., {Granot} J., {Paragi} Z., {Kouveliotou} C., {Wijers}
  R.~A.~M.~J., {Ramirez-Ruiz} E., 2010{\natexlab{a}}, GRB Coordinates Network,
  Circular Service, 11309, 1 (2010), 1309, 1

\bibitem[{{van der Horst} {et~al}\mbox{.}(2010{\natexlab{b}}){van der Horst},
  {Granot}, {Paragi}, {Kouveliotou}, {Wijers}, \&
  {Ramirez-Ruiz}}]{vanderhorstatel2874}
{van der Horst} A.~J., {Granot} J., {Paragi} Z., {Kouveliotou} C., {Wijers}
  R.~A.~M.~J., {Ramirez-Ruiz} E., 2010{\natexlab{b}}, The Astronomer's
  Telegram, 2874, 1

\bibitem[{{van der Horst} {et~al}\mbox{.}(2010{\natexlab{c}}){van der Horst},
  {Linford}, {Taylor}, {Paragi}, {Lundgren}, {de Ugarte Postigo}, {Belloni},
  {Kuulkers}, {Granot}, {Kouveliotou}, {Wijers}, \&
  {Garrett}}]{vanderhorstatel2918}
{van der Horst} A.~J. {et~al.}, 2010{\natexlab{c}}, The Astronomer's Telegram,
  2918, 1

\bibitem[{{van Paradijs} \& {McClintock}(1994)}]{vanparadijs1994}
{van Paradijs} J., {McClintock} J.~E., 1994, \aap, 290, 133

\bibitem[{{Wells}(1985)}]{wells1985}
{Wells} D.~C., 1985, in Data Analysis in Astronomy, {di Gesu} V., {Scarsi} L.,
  {Crane} P., {Friedman} J.~H., {Levialdi} S., eds., p. 195

\bibitem[{{Wijnands}, {Homan} \& {van der Klis}(1999){Wijnands}, {Homan}, \&
  {van der Klis}}]{wijnands1999}
{Wijnands} R., {Homan} J., {van der Klis} M., 1999, \apjl, 526, L33

\bibitem[{{Wilson} {et~al}\mbox{.}(2011){Wilson}, {Ferris}, {Axtens}, {Brown},
  {Davis}, {Hampson}, {Leach}, {Roberts}, {Saunders}, {Koribalski}, {Caswell},
  {Lenc}, {Stevens}, {Voronkov}, {Wieringa}, {Brooks}, {Edwards}, {Ekers},
  {Emonts}, {Hindson}, {Johnston}, {Maddison}, {Mahony}, {Malu}, {Massardi},
  {Mao}, {McConnell}, {Norris}, {Schnitzeler}, {Subrahmanyan}, {Urquhart},
  {Thompson}, \& {Wark}}]{wilson2011}
{Wilson} W.~E. {et~al.}, 2011, \mnras, 416, 832

\bibitem[{{Xu}(2010)}]{xugcn11303}
{Xu} D., 2010, GRB Coordinates Network, Circular Service, 11303, 1 (2010),
  1303, 1

\bibitem[{{Yamaoka} {et~al}\mbox{.}(2012){Yamaoka}, {Allured}, {Kaaret},
  {Kennea}, {Kawaguchi}, {Gandhi}, {Shaposhnikov}, {Ueda}, {Nakahira},
  {Kotani}, {Negoro}, {Takahashi}, {Yoshida}, {Kawai}, \&
  {Sugita}}]{yamaoka2012}
{Yamaoka} K. {et~al.}, 2012, \pasj, 64, 32

\end{thebibliography}

\newpage

\begin{center}
\begin{table}
\caption{UV, optical and nIR observations}
\label{tab:optical}
\renewcommand{\arraystretch}{1.1}
\begin{tabular}{|l|c|c|c|}
\hline
Epoch (MJD) & Telescope & Filter & Magnitude \\
\hline\hline
55468.050 & SMARTS & H & 14.70$\pm$0.13 \\
55468.995 & SMARTS & H & 14.85$\pm$0.12 \\
55469.999 & SMARTS & H & 14.88$\pm$0.21 \\
55470.990 & SMARTS & H & 14.90$\pm$0.18 \\
55472.002 & SMARTS & H & 15.02$\pm$0.08 \\
55472.985 & SMARTS & H & 15.22$\pm$0.15 \\
55476.002 & SMARTS & H & 15.33$\pm$0.14 \\
55476.994 & SMARTS & H & 15.60$\pm$0.14 \\
55478.010 & SMARTS & H & 15.40$\pm$0.14 \\
55478.997 & SMARTS & H & 15.50$\pm$0.18 \\
55479.997 & SMARTS & H & 15.49$\pm$0.14 \\
55481.006 & SMARTS & H & 15.54$\pm$0.24 \\
55483.002 & SMARTS & H & 15.51$\pm$0.21 \\
55468.044 & SMARTS & J & 15.13$\pm$0.13 \\
55468.990 & SMARTS & J & 15.27$\pm$0.19 \\
55469.993 & SMARTS & J & 15.26$\pm$0.10 \\
55470.984 & SMARTS & J & 15.32$\pm$0.21 \\
55471.996 & SMARTS & J & 15.26$\pm$0.16 \\
55472.979 & SMARTS & J & 15.41$\pm$0.23 \\
55475.996 & SMARTS & J & 15.52$\pm$0.11 \\
55476.988 & SMARTS & J & 15.84$\pm$0.21 \\
55478.004 & SMARTS & J & 15.68$\pm$0.11 \\
55478.991 & SMARTS & J & 15.76$\pm$0.10 \\
55479.991 & SMARTS & J & 15.70$\pm$0.10 \\
55481.001 & SMARTS & J & 15.83$\pm$0.10 \\
55482.997 & SMARTS & J & 15.79$\pm$0.06 \\
56497.977 & 3.5m CAHA & J & 21.05$\pm$0.17 \\
55710.964 & 2.2m CAHA &  z & 18.45$\pm$0.10 \\
55464.836 & IAC80 & I & 16.09$\pm$0.05 \\
55486.768 & 1.23m CAHA & I & 16.59$\pm$0.08 \\
55641.198 & 1.23m CAHA & I & 17.95$\pm$0.23 \\
55685.141 & 2.0m LT & i & 19.87$\pm$0.05 \\
55708.960 & 2.2m CAHA & i & 18.72$\pm$0.05 \\
55709.981 & 2.2m CAHA & i & 18.77$\pm$0.09 \\
55710.041 & 2.2m CAHA & i & 18.59$\pm$0.15 \\
55710.958 & 2.2m CAHA & i & 18.79$\pm$0.06 \\
55464.826 & IAC80 & R & 16.59$\pm$0.06 \\
55464.834 & IAC80 & R & 16.61$\pm$0.05 \\
55464.838 & BOOTES-2 & R & 16.59$\pm$0.09 \\
55464.838 & IAC80 & R & 16.58$\pm$0.06 \\
55464.839 & IAC80 & R & 16.62$\pm$0.05 \\
55464.840 & IAC80 & R & 16.59$\pm$0.06 \\
55464.840 & IAC80 & R & 16.56$\pm$0.06 \\
55464.841 & IAC80 & R & 16.58$\pm$0.06 \\
55464.841 & IAC80 & R & 16.56$\pm$0.06 \\
55464.842 & IAC80 & R & 16.59$\pm$0.06 \\
55464.842 & IAC80 & R & 16.57$\pm$0.06 \\
55464.843 & IAC80 & R & 16.55$\pm$0.06 \\
55464.843 & IAC80 & R & 16.56$\pm$0.07 \\
55464.846 & IAC80 & R & 16.57$\pm$0.07 \\
55464.849 & IAC80 & R & 16.54$\pm$0.07 \\
55466.799 & BOOTES-2 & R & 16.45$\pm$0.07 \\
55468.798 & BOOTES-2 & R & 16.41$\pm$0.06 \\
55469.790 & BOOTES-2 & R & 16.59$\pm$0.06 \\
55641.195 & 1.23m CAHA & R & 18.57$\pm$0.21 \\
55470.793 & BOOTES-2 & r & 16.69$\pm$0.20 \\
55474.358 & BOOTES-3 & r & 16.78$\pm$0.11 \\
\hline
\end{tabular}
\end{table}
\end{center}

\begin{center}
\begin{table}
\contcaption{UV, optical and nIR observations}
\renewcommand{\arraystretch}{1.1}
\begin{tabular}{|l|c|c|c|}
\hline
Epoch (MJD) & Telescope & Filter & Magnitude \\
\hline\hline
55474.383 & BOOTES-3 & r & 16.74$\pm$0.11 \\
55475.314 & BOOTES-3 & r & 17.01$\pm$0.11 \\
55475.336 & BOOTES-3 & r & 17.02$\pm$0.11 \\
55475.357 & BOOTES-3 & r & 16.98$\pm$0.11 \\
55475.378 & BOOTES-3 & r & 16.95$\pm$0.11 \\
55475.399 & BOOTES-3 & r & 16.84$\pm$0.11 \\
55708.960 & 2.2m CAHA & r & 19.03$\pm$0.04 \\
55709.981 & 2.2m CAHA & r & 19.09$\pm$0.04 \\
55710.041 & 2.2m CAHA & r & 19.15$\pm$0.12 \\
55710.961 & 2.2m CAHA & r & 19.05$\pm$0.05 \\
55464.627 & UVOT & V & 16.775$\pm$0.032 \\
55464.832 & IAC80 & V & 16.94$\pm$0.03 \\
55465.258 & UVOT & V & 16.676$\pm$0.032 \\
55465.735 & UVOT & V & 16.571$\pm$0.038 \\
55466.303 & UVOT & V & 16.525$\pm$0.031 \\
55467.635 & UVOT & V & 16.450$\pm$0.036 \\
55482.813 & UVOT & V & 17.197$\pm$0.083 \\
55483.489 & UVOT & V & 17.171$\pm$0.067 \\
55484.025 & UVOT & V & 17.211$\pm$0.080 \\
55485.028 & UVOT & V & 16.974$\pm$0.068 \\
55486.031 & UVOT & V & 17.046$\pm$0.068 \\
55487.036 & UVOT & V & 17.441$\pm$0.096 \\
55488.040 & UVOT & V & 17.178$\pm$0.079 \\
55489.044 & UVOT & V & 17.082$\pm$0.075 \\
55490.042 & UVOT & V & 17.337$\pm$0.107 \\
55491.045 & UVOT & V & 17.258$\pm$0.092 \\
55598.791 & UVOT & V & 18.576$\pm$0.284 \\
55620.663 & UVOT & V & $>$18.588 \\
55634.878 & UVOT & V & 18.756$\pm$0.303 \\
55641.195 & 1.23m CAHA & V & 18.67$\pm$0.21 \\
55647.369 & UVOT & V & $>$18.963 \\
55669.437 & UVOT & V & $>$18.798 \\
55687.980 & UVOT & V & 19.253$\pm$0.298 \\
55702.627 & UVOT & V & 18.562$\pm$0.214 \\
55717.735 & UVOT & V & $>$18.662 \\
55721.224 & UVOT & V & $>$18.642 \\
55725.361 & UVOT & V & $>$18.661 \\
55729.780 & UVOT & V & $>$18.622 \\
55733.057 & UVOT & V & $>$18.625 \\
55737.405 & UVOT & V & $>$18.788 \\
55741.027 & UVOT & V & $>$18.745 \\
55745.765 & UVOT & V & $>$18.648 \\
55749.584 & UVOT & V & $>$18.658 \\
55757.421 & UVOT & V & $>$18.610 \\
55759.752 & UVOT & V & $>$18.554 \\
55761.296 & UVOT & V & $>$18.344 \\
55470.793 & BOOTES-2 & g & 17.31$\pm$0.15 \\
55476.785 & BOOTES-2 & g & 17.42$\pm$0.17 \\
55708.960 & 2.2m CAHA & g & 19.60$\pm$0.07 \\
55709.981 & 2.2m CAHA & g & 19.66$\pm$0.05 \\
55710.041 & 2.2m CAHA & g & 19.81$\pm$0.19 \\
55710.961 & 2.2m CAHA & g & 19.62$\pm$0.07 \\
55464.666 & UVOT & B & 17.114$\pm$0.023 \\
55464.830 & IAC80 & B & 17.38$\pm$0.07 \\
55465.334 & UVOT & B & 17.014$\pm$0.026 \\
55465.727 & UVOT & B & 16.883$\pm$0.028 \\
55466.291 & UVOT & B & 16.818$\pm$0.023 \\
\hline
\end{tabular}
\end{table}
\end{center}

\begin{center}
\begin{table}
\contcaption{UV, optical and nIR observations}
\renewcommand{\arraystretch}{1.1}
\begin{tabular}{|l|c|c|c|}
\hline
Epoch (MJD) & Telescope & Filter & Magnitude \\
\hline\hline
55466.628 & UVOT & B & 16.804$\pm$0.030 \\
55467.629 & UVOT & B & 16.795$\pm$0.027 \\
55482.823 & UVOT & B & 17.378$\pm$0.046 \\
55483.502 & UVOT & B & 17.529$\pm$0.045 \\
55484.035 & UVOT & B & 17.482$\pm$0.049 \\
55485.038 & UVOT & B & 17.366$\pm$0.046 \\
55486.042 & UVOT & B & 17.460$\pm$0.048 \\
55487.046 & UVOT & B & 17.761$\pm$0.061 \\
55488.050 & UVOT & B & 17.483$\pm$0.050 \\
55489.054 & UVOT & B & 17.538$\pm$0.053 \\
55490.050 & UVOT & B & 17.769$\pm$0.072 \\
55491.055 & UVOT & B & 17.613$\pm$0.059 \\
55598.851 & UVOT & B & 18.908$\pm$0.128 \\
55620.657 & UVOT & B & 19.183$\pm$0.288 \\
55634.872 & UVOT & B & 19.434$\pm$0.240 \\
55641.198 & 1.23m CAHA & B & 19.71$\pm$0.48 \\
55647.363 & UVOT & B & 19.507$\pm$0.252 \\
55669.431 & UVOT & B & $>$19.745 \\
55687.975 & UVOT & B & 19.624$\pm$0.177 \\
55702.620 & UVOT & B & 19.281$\pm$0.176 \\
55717.730 & UVOT & B & $>$19.600 \\
55721.219 & UVOT & B & 19.440$\pm$0.342 \\
55725.356 & UVOT & B & $>$19.607 \\
55729.775 & UVOT & B & 19.073$\pm$0.256 \\
55733.052 & UVOT & B & $>$19.585 \\
55737.399 & UVOT & B & $>$19.738 \\
55741.021 & UVOT & B & $>$19.702 \\
55745.760 & UVOT & B & $>$19.605 \\
55749.579 & UVOT & B & $>$19.595 \\
55757.483 & UVOT & B & $>$19.963 \\
55759.748 & UVOT & B & $>$19.512 \\
55761.428 & UVOT & B & $>$19.893 \\
55464.662 & UVOT & U & 16.144$\pm$0.025 \\
55465.267 & UVOT & U & 16.036$\pm$0.025 \\
55465.725 & UVOT & U & 15.905$\pm$0.028 \\
55466.288 & UVOT & U & 15.845$\pm$0.025 \\
55466.625 & UVOT & U & 15.787$\pm$0.059 \\
55467.628 & UVOT & U & 15.816$\pm$0.027 \\
55468.494 & UVOT & U & 15.840$\pm$0.025 \\
55482.819 & UVOT & U & 16.450$\pm$0.037 \\
55483.496 & UVOT & U & 16.468$\pm$0.034 \\
55484.030 & UVOT & U & 16.565$\pm$0.039 \\
55485.033 & UVOT & U & 16.382$\pm$0.036 \\
55486.037 & UVOT & U & 16.510$\pm$0.037 \\
55487.041 & UVOT & U & 16.698$\pm$0.041 \\
55488.046 & UVOT & U & 16.550$\pm$0.039 \\
55489.049 & UVOT & U & 16.591$\pm$0.040 \\
55490.046 & UVOT & U & 16.725$\pm$0.047 \\
55491.050 & UVOT & U & 16.678$\pm$0.042 \\
55598.850 & UVOT & U & 18.624$\pm$0.132 \\
55620.656 & UVOT & U & 18.498$\pm$0.215 \\
55634.871 & UVOT & U & 18.469$\pm$0.146 \\
55647.362 & UVOT & U & 19.112$\pm$0.244 \\
55669.629 & UVOT & U & $>$19.660 \\
55687.974 & UVOT & U & 19.078$\pm$0.151 \\
55702.619 & UVOT & U & 18.715$\pm$0.149 \\
55717.729 & UVOT & U & 18.770$\pm$0.265 \\
\hline
\end{tabular}
\end{table}
\end{center}

\begin{center}
\begin{table}
\contcaption{UV, optical and nIR observations}
\renewcommand{\arraystretch}{1.1}
\begin{tabular}{|l|c|c|c|}
\hline
Epoch (MJD) & Telescope & Filter & Magnitude \\
\hline\hline
55721.218 & UVOT & U & 18.762$\pm$0.268 \\
55725.355 & UVOT & U & 18.685$\pm$0.242 \\
55729.774 & UVOT & U & 19.125$\pm$0.356 \\
55733.051 & UVOT & U & 19.132$\pm$0.351 \\
55737.397 & UVOT & U & 18.896$\pm$0.263 \\
55741.020 & UVOT & U & 18.591$\pm$0.216 \\
55745.759 & UVOT & U & 19.184$\pm$0.375 \\
55749.578 & UVOT & U & $>$19.275 \\
55757.482 & UVOT & U & $>$19.635 \\
55759.747 & UVOT & U & $>$19.202 \\
55761.427 & UVOT & U & $>$19.567 \\
55708.960 & 2.2m CAHA & u & 19.81$\pm$0.05 \\
55709.981 & 2.2m CAHA & u & 19.83$\pm$0.07 \\
55710.041 & 2.2m CAHA & u & 19.66$\pm$0.17 \\
55710.961 & 2.2m CAHA & u & 20.04$\pm$0.28 \\
55464.656 & UVOT & UVW1 & 16.438$\pm$0.024 \\
55465.260 & UVOT & UVW1 & 16.307$\pm$0.024 \\
55465.723 & UVOT & UVW1 & 16.196$\pm$0.027 \\
55466.284 & UVOT & UVW1 & 16.117$\pm$0.024 \\
55467.626 & UVOT & UVW1 & 16.103$\pm$0.026 \\
55471.384 & UVOT & UVW1 & 16.240$\pm$0.025 \\
55472.120 & UVOT & UVW1 & 16.317$\pm$0.027 \\
55475.400 & UVOT & UVW1 & 16.505$\pm$0.025 \\
55476.128 & UVOT & UVW1 & 16.483$\pm$0.029 \\
55476.664 & UVOT & UVW1 & 16.528$\pm$0.029 \\
55479.073 & UVOT & UVW1 & 16.697$\pm$0.035 \\
55479.541 & UVOT & UVW1 & 16.723$\pm$0.031 \\
55482.891 & UVOT & UVW1 & 16.639$\pm$0.047 \\
55484.576 & UVOT & UVW1 & 16.848$\pm$0.052 \\
55485.505 & UVOT & UVW1 & 16.823$\pm$0.053 \\
55488.578 & UVOT & UVW1 & 16.765$\pm$0.054 \\
55489.527 & UVOT & UVW1 & 16.993$\pm$0.063 \\
55490.586 & UVOT & UVW1 & 17.112$\pm$0.070 \\
55491.257 & UVOT & UVW1 & 17.165$\pm$0.071 \\
55598.848 & UVOT & UVW1 & 18.844$\pm$0.120 \\
55620.654 & UVOT & UVW1 & 19.022$\pm$0.257 \\
55634.869 & UVOT & UVW1 & 19.459$\pm$0.251 \\
55647.361 & UVOT & UVW1 & 19.150$\pm$0.208 \\
55669.628 & UVOT & UVW1 & $>$20.034 \\
55687.973 & UVOT & UVW1 & 19.722$\pm$0.204 \\
55702.616 & UVOT & UVW1 & 19.395$\pm$0.211 \\
55717.727 & UVOT & UVW1 & 19.349$\pm$0.341 \\
55721.216 & UVOT & UVW1 & 18.815$\pm$0.235 \\
55725.353 & UVOT & UVW1 & 19.126$\pm$0.284 \\
55729.773 & UVOT & UVW1 & 19.286$\pm$0.325 \\
55733.050 & UVOT & UVW1 & 19.360$\pm$0.353 \\
55737.396 & UVOT & UVW1 & $>$19.658 \\
55741.019 & UVOT & UVW1 & $>$19.614 \\
55745.758 & UVOT & UVW1 & $>$19.419 \\
55749.577 & UVOT & UVW1 & $>$19.490 \\
55757.481 & UVOT & UVW1 & $>$19.905 \\
55759.746 & UVOT & UVW1 & $>$19.412 \\
55761.426 & UVOT & UVW1 & $>$19.846 \\
55464.623 & UVOT & UVW2 & 16.548$\pm$0.028 \\
55465.324 & UVOT & UVW2 & 16.470$\pm$0.028 \\
55465.732 & UVOT & UVW2 & 16.326$\pm$0.025 \\
55466.297 & UVOT & UVW2 & 16.251$\pm$0.023 \\
\hline
\end{tabular}
\end{table}
\end{center}

\begin{center}
\begin{table}
\contcaption{UV, optical and nIR observations}
\renewcommand{\arraystretch}{1.1}
\begin{tabular}{|l|c|c|c|}
\hline
Epoch (MJD) & Telescope & Filter & Magnitude \\
\hline\hline
55467.632 & UVOT & UVW2 & 16.245$\pm$0.024 \\
55469.478 & UVOT & UVW2 & 16.399$\pm$0.026 \\
55473.392 & UVOT & UVW2 & 16.506$\pm$0.025 \\
55477.132 & UVOT & UVW2 & 16.602$\pm$0.030 \\
55477.668 & UVOT & UVW2 & 16.819$\pm$0.031 \\
55482.880 & UVOT & UVW2 & 16.853$\pm$0.050 \\
55484.565 & UVOT & UVW2 & 17.014$\pm$0.052 \\
55485.494 & UVOT & UVW2 & 17.094$\pm$0.055 \\
55488.569 & UVOT & UVW2 & 17.111$\pm$0.057 \\
55489.519 & UVOT & UVW2 & 17.095$\pm$0.060 \\
55490.579 & UVOT & UVW2 & 17.366$\pm$0.072 \\
55491.249 & UVOT & UVW2 & 17.348$\pm$0.069 \\
55598.852 & UVOT & UVW2 & 19.249$\pm$0.100 \\
55620.660 & UVOT & UVW2 & 19.859$\pm$0.302 \\
55634.875 & UVOT & UVW2 & 19.889$\pm$0.214 \\
55647.366 & UVOT & UVW2 & 19.751$\pm$0.201 \\
55669.434 & UVOT & UVW2 & $>$20.301 \\
55687.977 & UVOT & UVW2 & 20.305$\pm$0.201 \\
55702.624 & UVOT & UVW2 & 19.505$\pm$0.143 \\
55717.732 & UVOT & UVW2 & 19.880$\pm$0.310 \\
55721.222 & UVOT & UVW2 & 19.441$\pm$0.231 \\
55725.358 & UVOT & UVW2 & 19.291$\pm$0.201 \\
55729.778 & UVOT & UVW2 & 19.463$\pm$0.232 \\
55733.055 & UVOT & UVW2 & 19.332$\pm$0.216 \\
55737.402 & UVOT & UVW2 & 19.624$\pm$0.235 \\
55741.024 & UVOT & UVW2 & 19.672$\pm$0.250 \\
55745.763 & UVOT & UVW2 & 19.993$\pm$0.344 \\
55749.582 & UVOT & UVW2 & $>$20.115 \\
55757.484 & UVOT & UVW2 & $>$20.230 \\
55759.750 & UVOT & UVW2 & $>$20.005 \\
55761.430 & UVOT & UVW2 & $>$20.298 \\
55464.633 & UVOT & UVM2 & 16.950$\pm$0.033 \\
55465.250 & UVOT & UVM2 & 16.835$\pm$0.029 \\
55465.739 & UVOT & UVM2 & 16.687$\pm$0.032 \\
55466.270 & UVOT & UVM2 & 16.683$\pm$0.033 \\
55467.638 & UVOT & UVM2 & 16.602$\pm$0.032 \\
55470.481 & UVOT & UVM2 & 16.702$\pm$0.031 \\
55474.362 & UVOT & UVM2 & 17.023$\pm$0.032 \\
55478.136 & UVOT & UVM2 & 17.174$\pm$0.042 \\
55478.671 & UVOT & UVM2 & 17.202$\pm$0.042 \\
55482.886 & UVOT & UVM2 & 17.228$\pm$0.072 \\
55484.571 & UVOT & UVM2 & 17.435$\pm$0.077 \\
55485.500 & UVOT & UVM2 & 17.519$\pm$0.082 \\
55488.574 & UVOT & UVM2 & 17.511$\pm$0.084 \\
55489.523 & UVOT & UVM2 & 17.451$\pm$0.087 \\
55490.583 & UVOT & UVM2 & 17.760$\pm$0.107 \\
55491.253 & UVOT & UVM2 & 17.674$\pm$0.100 \\
55598.794 & UVOT & UVM2 & 19.738$\pm$0.296 \\
55620.666 & UVOT & UVM2 & $>$19.757 \\
55634.880 & UVOT & UVM2 & 20.274$\pm$0.374 \\
55647.370 & UVOT & UVM2 & $>$20.039 \\
55669.440 & UVOT & UVM2 & $>$19.817 \\
55687.982 & UVOT & UVM2 & $>$20.656 \\
55702.630 & UVOT & UVM2 & 19.713$\pm$0.240 \\
55717.737 & UVOT & UVM2 & $>$19.584 \\
55721.226 & UVOT & UVM2 & $>$19.728 \\
55725.363 & UVOT & UVM2 & $>$19.763 \\
\hline
\end{tabular}
\end{table}
\end{center}

\begin{center}
\begin{table}
\contcaption{UV, optical and nIR observations}
\renewcommand{\arraystretch}{1.1}
\begin{tabular}{|l|c|c|c|}
\hline
Epoch (MJD) & Telescope & Filter & Magnitude \\
\hline\hline
55729.782 & UVOT & UVM2 & $>$19.587 \\
55733.059 & UVOT & UVM2 & $>$19.648 \\
55737.407 & UVOT & UVM2 & $>$19.720 \\
55741.029 & UVOT & UVM2 & 19.780$\pm$0.369 \\
55745.767 & UVOT & UVM2 & $>$19.724 \\
55749.586 & UVOT & UVM2 & $>$19.752 \\
55757.423 & UVOT & UVM2 & $>$19.720 \\
55759.754 & UVOT & UVM2 & $>$19.625 \\
55761.298 & UVOT & UVM2 & $>$19.405 \\
\hline
\end{tabular}
\end{table}
\end{center}

\end{document}